\documentclass[8.5pt,twoside,twocolumn]{article}
\usepackage[super,sort&compress,comma]{natbib} 
\usepackage{mhchem}
\usepackage{times,mathptmx}
\usepackage{sectsty}
\usepackage{balance} 

\usepackage{graphicx} 
\usepackage{lastpage}
\usepackage[format=plain,singlelinecheck=false,font=small,labelfont=bf,labelsep=space]{caption} 
\usepackage{fancyhdr}
\usepackage{epsfig,amsmath,amsbsy,amssymb}

\begin{document}

\thispagestyle{plain}
\fancypagestyle{plain}{
\renewcommand{\headrulewidth}{1pt}}
\renewcommand{\thefootnote}{\fnsymbol{footnote}}
\renewcommand\footnoterule{\vspace*{1pt}%
\hrule width 3.4in height 0.4pt \vspace*{5pt}} 
\setcounter{secnumdepth}{5}

\makeatletter 
\def\subsubsection{\@startsection{subsubsection}{3}{10pt}{-1.25ex plus -1ex minus -.1ex}{0ex plus 0ex}{\normalsize\bf}} 
\def\paragraph{\@startsection{paragraph}{4}{10pt}{-1.25ex plus -1ex minus -.1ex}{0ex plus 0ex}{\normalsize\textit}} 
\renewcommand\@biblabel[1]{#1}            
\renewcommand\@makefntext[1]%
{\noindent\makebox[0pt][r]{\@thefnmark\,}#1}
\makeatother 
\renewcommand{\figurename}{\small{Fig.}~}
\sectionfont{\large}
\subsectionfont{\normalsize} 

\fancyfoot{}
\fancyhead{}
\renewcommand{\headrulewidth}{1pt} 
\renewcommand{\footrulewidth}{1pt}
\setlength{\arrayrulewidth}{1pt}
\setlength{\columnsep}{6.5mm}
\setlength\bibsep{1pt}

\twocolumn[
  \begin{@twocolumnfalse}
\noindent\LARGE{\textbf{Semiflexible polymer rings on topographically and chemically  structured surfaces}}
\vspace{0.6cm}

\noindent\large{\textbf{Petra Gutjahr,\textit{$^{a}$} Reinhard Lipowsky\textit{$^{a}$} and
Jan Kierfeld\textit{$^{ab}$}}}\vspace{0.5cm}

\noindent\textit{\small{\textbf{Received 14th May 2010, Accepted 23rd August 2010
}}}

\noindent \textbf{\small{DOI: 10.1039/c0sm00381f}}
\vspace{0.6cm}

\noindent \normalsize{We investigate morphologies of 
 semiflexible polymer rings, such as circular DNA, which  are adsorbed
 onto  topographically or chemically structured substrate surfaces.
 We classify all equilibrium morphologies for two striped surface 
 structures, (i)
  topographical surface grooves  and (ii)   chemically structured 
   surface domains.
 For both types of stripes, we 
 find four equilibrium shapes: a  round toroidal and  a confined elongated
shape as well as two shapes containing bulges.
 We determine the complete 
 bifurcation diagram of these morphologies
 as a function of  their contour length and the ratio of adhesive
 strength  to bending rigidity.
 For more  complex geometries consisting of several stripes we find a
 cascade of transitions between elongated shapes. 
 Finally, we compare our findings to ring  condensation by 
 attractive interactions.}
\vspace{0.5cm}
 \end{@twocolumnfalse}
  ]

 \section{Introduction}
\footnotetext{\dag~Electronic Supplementary Information (ESI) available: [details of any supplementary information available should be included here]. See DOI: 10.1039/b000000x/}


\footnotetext{\textit{$^{a}$~Max Planck Institute of Colloids and Interfaces, Science
   Park Golm, 14424 Potsdam, Germany}}
\footnotetext{\textit{$^{b}$~Physics Department, TU Dortmund University,
   44221 Dortmund, Germany}}


\footnotetext{\ddag~Additional footnotes to the title and authors can be included \emph{e.g.}\ `Present address:' or `These authors contributed equally to this work' as above using the symbols: \ddag, \textsection, and \P. Please place the appropriate symbol next to the author's name and include a \texttt{\textbackslash footnotetext} entry in the the correct place in the list.}

 Bionanotechnology requires the immobilization and controlled
 manipulation of DNA  and other semiflexible polymers. 
 Adsorption is the simplest technique  to immobilize 
 single polymers and a first step towards further 
 visualization and manipulation using, e.g., modern  scanning probe
 techniques \cite{SM01,Sam04}. For manipulation, control over the 
 shape  of the adsorbed  polymer is needed. 
 In this article,  we explore the possibility to achieve such
 shape control for   semiflexible polymer rings  using 
 simple striped surface structures, which can be realized
 by topographical or chemical structuring. 
 Whereas flexible polymers are governed by
 conformational entropy and typically adsorb in  random
 coil configurations, 
 the morphologies of  semiflexible polymers with large
  persistence lengths  are dominated by their bending rigidity, 
which gives rise to well-defined shapes: An open 
 polymer adsorbs in a straight configuration, whereas a closed
  polymer ring forms a circular loop. Examples of such semiflexible rings
 are provided by 
 DNA minicircles \cite{Mad06}, carbon nanotubes \cite{Sano01}
 filamentous actin  \cite{Tang01}, and  amyloid fibrils \cite{Hat03}.
The shape of such semiflexible rings is of importance 
for biological issues such as the accessibility 
in transcription of viral genomes or plasmids or their 
transport properties.

 In the absence of thermal fluctuations or external forces 
closed semiflexible rings assume a well-defined circular 
shape. It has been shown that thermal fluctuations lead to 
interesting  effects such as a gradual  crossover from oblate 
spherically symmetric 
ring shapes at low temperatures or high stiffness to 
prolate shapes for increasingly flexible rings with 
a maximal asphericity at intermediate stiffnesses \cite{AF07}.
Semiflexible rings may also  be viewed as one-dimensional analogues 
of two-dimensional vesicles, for which  thermal fluctuations 
gives rise to shape asphericity as well \cite{LLG05}.
In this article, we focus on 
the influence of external forces or potentials, which tend to confine 
the semiflexible ring. This issue is of general importance not only 
to control the shape of semiflexible 
polymers for further manipulation but also to understand  how 
semiflexible polymers or elastic sheets can be 
packaged  and 
which forces or potentials 
are necessary to achieve a given packaging configuration.

\begin{figure*}
 \begin{center}
  \includegraphics[width=0.99\textwidth]{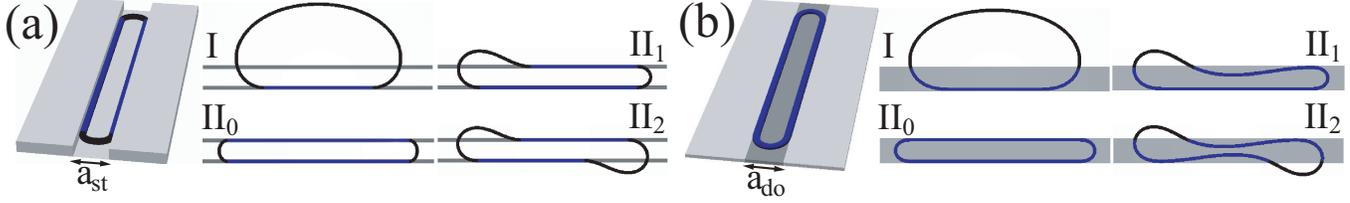}
 \caption{\label{types}  
 Adsorbed polymer on a striped surface containing 
  (a) a rectangular 
 topographical surface groove of width $a_{st}$ 
 and (b) a chemically structured 
  surface domain of width $a_{do}$. 
Both in (a) and (b), the first perspective drawings illustrate the 
system geometry whereas the remaining four subfigures represent 
 top views of all  four  stable ring morphologies ${\rm I}$,  ${\rm II_1}$, 
${\rm II_0}$, and ${\rm II_2}$ as obtained by energy
 minimization for contour lengths $L/a_{st}=20$ and  $L/a_{do}=20$;
 the perspective drawings correspond to the elongated shape ${\rm II_0}$.  
 }
 \end{center}
 \end{figure*}

 In this article, we present a quantitative analysis of 
  shapes  of a strongly adsorbed  semiflexible polymer
 ring in the  presence of an additional substrate structure,
  which can be  either a topographical surface groove with rectangular 
cross section, 
  see Fig.\ \ref{types}(a),  or a  
   striped domain of increased adhesion energy, see Fig.\
  \ref{types}(b).
  Topographical surface steps have been employed  in recent manipulation 
  experiments on semiflexible polymer rings \cite{rabe}.
 Adsorption of DNA on  grooved, periodically structured 
substrates has also been investigated  in Refs.\
\citenum{hochrein06,hochrein07}. 
  Both striped structures
   introduce a laterally modulated adhesion potential,
  which attracts the ring to the stripe.
 We find that, for persistence lengths larger than the stripe width, 
 the   competition
  between its bending rigidity and the attraction to the striped 
  domain allows a  controlled switching
 between four distinct stable  morphologies,  see Fig.\ \ref{types}. 
 Apart from a weakly bound almost circular shape
  and a strongly bound elongated shape,
  bulged intermediate shapes become stable for large
  contour lengths. We determine the full 
 bifurcation diagrams for  semiflexible ring  shapes 
 both analytically and numerically. 
 This analysis
 can  be used to (i) control the ring shape and (ii)
 analyze material properties of the
 substrate or the semiflexible polymer ring experimentally.
 Flexible  polymer rings, on the other hand, 
  exhibit random coil configurations and do not undergo 
  such morphological transitions.
 Finally, we  generalize our findings to  semiflexible polymer rings on 
   a {\em periodic} stripe pattern, 
  which serves as a model for the interaction between
  a  polymer and the atomic lattice structure of substrate surface.

 The condensation of  semiflexible polymers 
 such as DNA \cite{bloomfield91,bloomfield97,schnurr2000,schnurr2002}
  or actin filaments
 \cite{Tang01,Cebers06} is a  closely related  transition phenomenon 
 that is caused by the competition between attractive interactions
and bending energy.
 In poor solvent, in the presence of condensing agents or depletion forces,
   polymer-polymer contacts become favorable, 
  but the bending rigidity of a semiflexible polymer inhibits its
  collapse towards a tightly packed globular structure, which is
   common for flexible polymers.
 As a result, semiflexible polymers form toroidal bundles
  \cite{bloomfield97,Tang01,Cebers06}  %
   via a cascade of metastable racquet states \cite{schnurr2000}. 
 At the end of the article, we compare our findings for 
 morphological transitions on structured substrates 
 to the condensation of a semiflexible ring 
 by attractive polymer-polymer interactions.

 \section{Topographical surface groove}

\subsection{Model}

 We consider a  semiflexible polymer ring of fixed contour length $L$
 adsorbed to a planar substrate in the $xy$-plane that  contains two parallel
 topographical surface steps at $x=\pm a_{st}/2$
  forming an infinitely long rectangular  surface groove
 of width $a_{st}$, see Fig.\  \ref{types}(a). 
 It is  further assumed that the
 overall adhesion is so strong, that the polymer is firmly adsorbed
 to the substrate surface. We will first focus on 
  polymer morphologies at zero 
 temperature $T=0$ and discuss the effect of thermal
  fluctuations in the end. The polymer gains an
 additional adsorption energy $W_{st}<0$ per polymer length only at
 the corners of the rectangular 
surface groove, where it  can bind to two adjacent
 surfaces as shown in Fig.\  \ref{types}(a). 
 The resulting lateral adsorption potential can be
 described by $V_{st}(x) = W_{st}$ for $|x\pm a_{st}/2|<\ell/2$  and 
 $V_{st}(x) = 0$
 otherwise, where $\ell$ denotes the adhesive range of the surface steps, 
 which is of the order of the polymer diameter
 and  assumed to be small 
 compared to the groove width, $\ell \ll a_{st}$.

 The  bending energy of the polymer is given by
 \begin{equation}
  E_b = (\kappa/2)\int_0^{L} ds (\partial_s \theta(s))^2,
\label{Eb}
\end{equation}
 where $\kappa$ is the bending rigidity, and the contour is
 parameterized by the arc length $s$ ($0<s<L$) using  the tangent
 angles $\theta(s)$. 
 The adhered length $L_{st}$ is given by the polymer length
 on the edges at $x=\pm a_{st}/2$, and the adhesion energy is
 \begin{equation}
  E_{ad} = -|W_{st}|L_{st}.
\label{Ead}
\end{equation}
 In the following we often use dimensionless 
 quantities by  measuring 
 lengths in units of the groove width, 
 and  energies 
  in units of the typical bending energy, 
\begin{equation}
   \bar{L}\equiv L/a_{st} ~~\mbox{and}~~\bar{E} \equiv Ea_{st}/\kappa.
\label{barE}
\end{equation}
This leads to $\bar{E}_{ad} = -|w_{st}|\bar{L}_{st}$
with a {\em reduced adhesion strength} 
\begin{equation}
  |w_{st}| \equiv
|W_{st}|a_{st}^2/\kappa.
\label{wst}
\end{equation}

The polymer configuration is determined by minimizing
 the total energy $E_{\rm tot}=E_b+E_{ad}$ 
 under the constraints imposed by  ring closure, i.e.,
 $\int_0^{L} ds (\cos{\theta(s)},\sin{\theta(s)}) = (0,0)$.
 This yields a shape equation
 for $\theta(s)$ and an implicit equation for the Lagrange multiplier
 associated with the ring closure constraint.
 Solving these equations, the polymer shape and the resulting
  energies can be  calculated analytically.
 
We assume the surface step height to be comparable to the polymer
  diameter and  neglect small energy corrections arising if the polymer
 crosses the surface steps. 
For large step heights and large bending rigidity $\kappa$ 
these corrections become important, and 
a polymer crossing two parallel
 topographical surface steps can even lift from 
the groove in this limit \cite{pierrelouis08}.

\subsection{Energy minima for fixed adhered length}

 For a single value of the adhesion strength $W_{st}$ we  often find 
several metastable ring shapes apart from the ring shape representing the
stable global energy minimum. 
 In order to determine  all  metastable states 
 of the total energy $E_{\rm tot}= E_b+E_{ad}$ and discuss bifurcations
of these metastable states, 
 we   minimize the  {\em constrained}  energy $E_{\rm tot}(L_{st})$,
 where we also fix  the  adhered length $L_{st}$ and, thus, the 
 adhesion energy. Then, each metastable state represents a local minimum 
in the energy landscape given by the function $E_{\rm tot}(L_{st})$. 
If a local minimum vanishes, the  corresponding metastable state 
becomes unstable. If the global minimum exchanges between two 
local minima, we are at a transition point between two 
morphologies. 
Maxima in the constrained energy  also allow
 us to discuss possible transition states of these 
shape transitions.

For a topographical groove all metastable states consist of one or two 
 straight adhered segments at  the corners of the 
groove  and with total length
 $L_{st}$, which are connected by one or two curved segments, respectively.
The curved segments are bending energy minimizers, i.e., planar 
Euler elastica. 
 More than two curved segments are unfavorable.
 We find four possible  metastable morphologies: 
 For small $L_{st}$, the ring
 will adhere only to one corner of the groove 
and adopt the rather round toroidal
 configuration ${\rm I}$, see Fig.~\ref{types}. 
 For $L_{st} \gtrsim L/2$,  conformations of the type ${\rm II}$, 
 where the ring binds 
 to both corners, will become relevant.
 These shapes consist of two round segments connecting straight adhered
 segments. The round  segments can have either the form of a round {\em cap} 
 staying completely inside the groove or contain {\em bulges}, which 
are round segments outside the groove.
All shapes adhering to both  corners of the groove
  may be classified by the number of bulges 
and are referred to as ${\rm II_0}$,
 ${\rm II_1}$, and ${\rm II_2}$,  accordingly, see Fig.~\ref{types}. 

 In order to minimize  bending energy, 
  curved segments in shapes of type ${\rm II}$ will only bulge to 
 one side of the groove. 
In principle, also  curved segments extending  bulges 
to both sides of the groove resulting in a configuration with 
with reflection symmetry with respect to the $x$-axis represent 
a metastable state.
As derived in the appendix, the bending energy of an 
asymmetric one-sided bulge is lower in bending energy for the 
same adhered length $L_{st}$ for {\em all} 
possible adhered lengths. 
Therefore, we neglect this type of bulges in the following and 
 consider  only 
the remaining  four relevant metastable 
 shapes ${\rm I}$ , ${\rm II_0}$,
 ${\rm II_1}$, and ${\rm II_2}$.

The curved segments on both sides of a ring in  shapes ${\rm II}$ 
can exchange length 
even if the total adhered length $L_{st}$ is fixed. This results in the 
transversality condition that the curvatures at  the contact points 
have to be equal in all four contact points of shapes ${\rm II}$. 
Therefore, bulges or caps on both sides of the ring 
have to have the same size in shapes ${\rm II_0}$ and ${\rm II_2}$.
In shape ${\rm II_2}$ bulges at both ends can be on either side of the 
groove, therefore two energetically degenerate shapes ${\rm II_2}$ 
exist for a topographical groove. The shape with both bulges 
on the same side has reflection symmetry with respect to the $y$-axis, 
the shape with both bulges on opposite sides, which is shown 
in Fig.~\ref{types}(a), is antisymmetric
with respect to the center of the shape.

\begin{figure}
\begin{center}
  \includegraphics[width=0.46\textwidth]{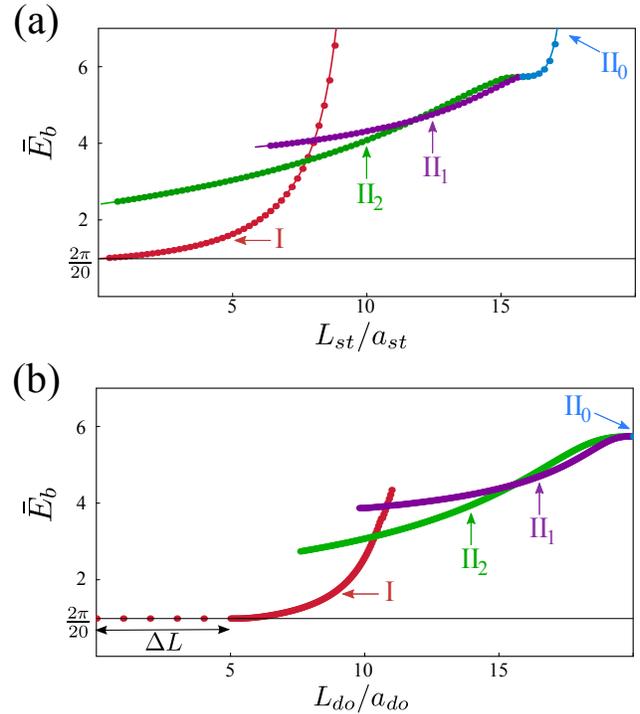}
\caption{\label{fig:landscape}
(a) The dimensionless 
 bending energy $\bar{E}_{b}$ as defined in (\ref{barE}) 
of the  metastable states ${\rm I}$ (red), ${\rm II_0}$ (blue), 
 ${\rm II_1}$ (violet), and ${\rm II_2}$ (green),  as a function of
the adhered length $L_{st}/a_{st}$ 
for a semiflexible ring of length $L/a_{st} = 20$ adsorbing on a
{\em topographical surface groove} of width $a_{st}$.
Solid lines are analytical results, dots represent data from numerical energy
minimization. The arrows correspond to the shapes shown in Fig.\
\ref{types}(a). 
(b) The  dimensionless
bending energy $\bar{E}_b$ as defined in (\ref{barE2})
of the  metastable states as a function of the adhered length 
$L_{do}$  (in units of $a_{do}$) for a semiflexible ring of 
length $L/a_{do} = 20$  adhering  to a
{\em chemically structured stripe} of width $a_{do}$
from numerical energy minimization.
There is an approximately  constant shift in the adhered length
$\Delta \bar{L}$ compared to the results for the topographical stripe in (a).
The
arrows point to shapes that are displayed in Fig.\ \ref{types}(b).
}
\end{center}
\end{figure}
 
For a topographical stripe the energies of the
four types of shapes can be obtained by 
 analytical energy minimization. 
For fixed adhered length $L_{st}$ the 
adhesion energy  $\bar{E}_{ad} = - |w_{st}|\bar{L}_{st}$
is a constant contribution to the constrained energy $E_{\rm tot}(L_{st})$.
Therefore, 
the analytical calculation starts by calculating  the bending energy 
$\bar{E}_b(\bar{L}_{st})$  of all four metastable states 
as a function of the adhered length $\bar{L}_{st}$.
This bending energy arises from the curved segments of the 
ring shape, which take the shape of Euler elastica. 
To calculate these shapes a constraint for ring closure has 
to be imposed, which is associated with an additional Lagrange 
multiplier. For the resulting second order 
Euler Lagrange equations describing the 
shapes of curved segments can always find one first integral.
We can express contour length $L$, adhered length $L_{st}$ and 
bending energy $E_b$ parametrically as functions of a 
single parameter 
which is  related to the integration constant of the 
shape equations.  
Using this method, we derive analytical parametric representations of the 
bending energy landscape $\bar{E}_b(\bar{L}_{st})$
in terms of this integration constant parameter  in the appendix. 
 In  Fig.\ \ref{fig:landscape}(a) 
we show the main result of this calculation, which is 
 the bending energy landscape 
$\bar{E}_b(\bar{L}_{st})$ consisting of four bending energy branches 
corresponding to the four different morphologies.

Our analytical results are confirmed by numerical energy minimization 
using  the dynamical discretization algorithm of the SURFACE EVOLVER
2.14~\cite{brakke}. 
In Fig.\  \ref{fig:landscape}(a),
 the bending energy landscapes $\bar{E}_{b}(\bar{L}_{st})$
of all four  morphologies  as obtained from the exact
analytical energy minimization are shown as solid lines; the results 
from numerical  energy minimization as colored dots and 
 and completely agree with the analytical results.

\subsection{Unconstrained energy minima}

\begin{figure}
\begin{center}
 \includegraphics[width=0.46\textwidth]{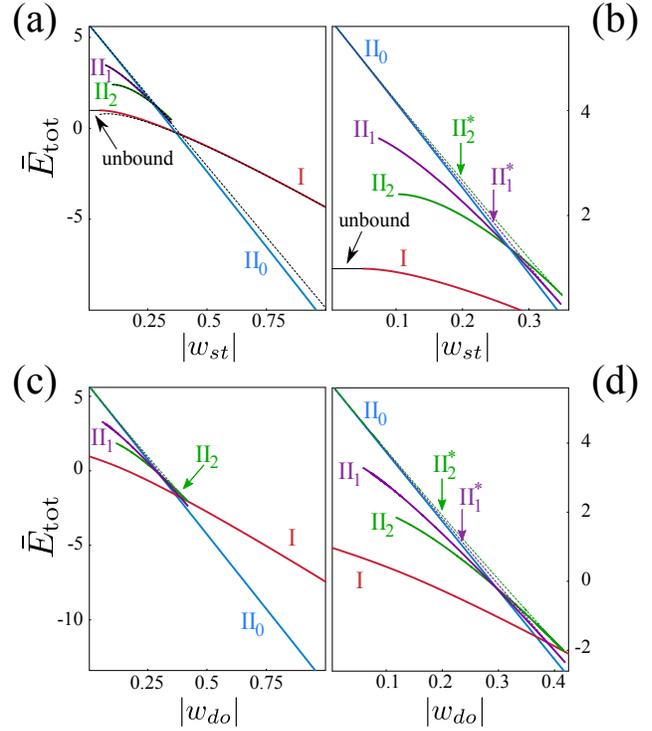}
 \caption{\label{eminv} 
(a) The dimensionless total energy $\bar{E}_{\rm tot}$ of 
all metastable states
versus the reduced potential $|w_{st}|$ for
a semiflexible ring of contour length $L/a_{st} = 20$ adsorbing on a  
{\em  topographical surface groove} of width $a_{st}$. 
The graph shows  analytical  results for the 
  shapes  ${\rm I}$, ${\rm II_0}$,
 ${\rm II_1}$, and ${\rm II_2}$  as
 red, blue, violet, and green solid lines, respectively. 
Analytical estimates for these energy curves as derived in the appendix 
are shown  as dashed lines.
Lines end where a metastable shape become unstable. 
(b) A magnification of the upper left corner of (a) including 
the unstable transition states ${\rm II_1^*}$ and ${\rm II_2^*}$ as violet
and green dashed line,
respectively. The analytical  estimates are omitted for clarity.
(c) Analogous numerical results of the dimensionless total
 energy $\bar{E}_{\rm tot}$
as a function of $|w_{do}|$
for a semiflexible ring adhered to a {\em chemically structured stripe}
  with $L/a_{do} = 20$ and magnified in (d).
The unbound circle is never stable and therefore absent 
for the chemical domain.
 }
\end{center}
\end{figure}

The  total 
energy landscape  $\bar{E}_{\rm tot}(\bar{L}_{st},|w_{st}|)$ 
for each morphology 
is obtained by adding the linearly decreasing 
 adhesion energy  $\bar{E}_{ad} = - |w_{st}|\bar{L}_{st}$ to the corresponding 
bending energy branch, i.e., by a simple 
 tilt of the bending energy landscape in Fig.\  \ref{fig:landscape}(a).
The local minimum of each branch of the resulting energy landscape 
 $\bar{E}_{\rm tot}(\bar{L}_{st},|w_{st}|)$ with respect to the 
adhered length $(\bar{L}_{st}$ gives the 
equilibrium total energy of the corresponding morphology 
${\rm I}$, ${\rm II_{0}}$, ${\rm II_{1}}$, or  ${\rm II_{2}}$.
These energy minima depend on the tilt of the bending energy 
landscape and are, thus,  a function of the 
adhesion strength $|w_{st}|$,
\begin{equation}
  \bar{E}_{\rm tot}(|w_{st}|) =
    \min_{\bar{L}_{st}} \left( \bar{E}_{b}(\bar{L}_{st})
          -|w_{st}|\bar{L}_{st}\right)
\label{Legendre}
\end{equation}
Therefore, the equilibrium total energy $\bar{E}_{\rm tot}(|w_{st}|)$
is a Legendre transform  of the bending energy $\bar{E}_{b}(\bar{L}_{st})$ 
with respect to the adhered length $\bar{L}_{st}$. 
The  resulting 
energy bifurcation diagram Fig.~\ref{eminv}(a,b)  shows the 
four branches of the 
equilibrium total energy $\bar{E}_{\rm tot}$  for  the four  local minima
as a function of $|w_{st}|$ for a contour length $\bar{L}=20$.

If the constraint on the total adhered length is lifted and the total 
energy minimized with respect to the 
adhered length, a transversality condition arises 
at the contact points
where the curved segments  join the  
straight segments adhered on the surface steps. 
According to this  transversality condition
the   curvature at the contact points  is given by   
the inverse contact radius \cite{seifertlipowsky1990,seifert1991}
\begin{equation}
   1/R_{co} = (2|W_{st}|/\kappa)^{1/2}.
\label{Rco}
\end{equation}
In the limit of small bending rigidity $\kappa$ and  large adhesion 
strength $|W_{st}|$  the contact radius becomes small compared to the 
 groove  width,  $R_{co} \ll a_{st}$, corresponding to $|w_{st}| \gg 1$
for the reduced adhesion strength (\ref{wst}).
In this limit, the ring assumes an effectively kinked shape
${\rm II_0}$,
in which caps on both sides of the ring become almost straight and 
the ring shape resembles a rectangle 
with sharp kinks, similar to shapes 
that have been observed in Refs.\  \citenum{hochrein06,hochrein07}.
Only on  length scales smaller than the contact radius $R_{co}$ these 
sharp kinks can be resolved as smooth bends. 

In the appendix we  present exact analytical results for the 
unconstrained total energy $\bar{E}_{\rm tot}(|w_{st}|)$:
Using the condition of contact curvature at the end points 
of all curved segments, we 
 derive analytical parametric representations of the 
total  energy $\bar{E}_{\rm tot}(|w_{st}|)$
in terms of the same  integration constant parameter 
which we used for the bending energy landscapes 
$\bar{E}_b(\bar{L}_{st})$. 
In the following, we focus on 
approximate results for the total energy and 
outline the main features of the bifurcation diagram.

In the bifurcation diagram Fig.~\ref{eminv}(a,b), 
the globally stable ring shape  is the shape with the lowest 
energy $\bar{E}_{\rm tot}$  for a given adhesion strength 
 $|w_{st}|$.
 If two branches of local minima in the bifurcation diagrams 
in Fig.~\ref{eminv}  cross or merge, a morphological transition between 
the corresponding shapes occurs. If the branches cross at a finite angle, 
this transition is   discontinuous with 
hysteresis effects and 
 jumps in the adhered length $L_{st}$, which is given by 
the negative slope 
$L_{st} = -\partial_{|w_{st}|}\bar{E}_{\rm tot}(|w_{st}|)$
according to (\ref{Legendre}). 
As can be seen from the bifurcation diagram  Fig.~\ref{eminv}(a,b) 
 all shape transition 
between the four metastable states are {\em discontinuous}, 
which gives rise to many metastable shapes and shape hysteresis. 
A metastable state becomes unstable if the corresponding branch ends. 
In particular, this happens for the two bulged shapes ${\rm II_1}$ and 
${\rm II_2}$, which are only metastable for a limited range of 
adhesion strengths $|w_{st}|$.

Apart from discontinuous transitions between the four shapes,
shape ${\rm I}$ undergoes 
 an additional {\em continuous} unbinding transition from a single 
surface step, 
which is also known for  vesicles adhering to a
surface, where  the interplay between
adhesion and bending energy 
 leads to an unbinding transition, which is not driven by
 thermal fluctuations \cite{seifertlipowsky1990,seifert1991}.
A transversality condition enforces the
curvature at the contact point 
to be $1/R_{co}=(2|w_{st}|)^{1/2}/a_{st}$, see eq.\ (\ref{Rco}), 
 such that 
rings of contour length $L$ can only bind to the contact line for 
   $\bar{L} \ge  \bar{L}_{ub}$ with 
\begin{equation}
   \bar{L}_{ub} = \sqrt{2} \pi |w_{st}|^{-1/2}.
\label{unbinding}
\end{equation}

In the energy  bifurcation diagrams Fig.~\ref{eminv}(a,b)
the round configuration ${\rm I}$ (red line) represents the global
energy minimum  for small $|w_{st}|$, whereas the adhesion energy gain
dominates for large $|w_{st}|$, and 
 the elongated shape ${\rm II_0}$ 
(blue line) becomes the globally stable conformation.
For  adhesion strengths $|w_{st}|\bar{L}^{2} \gg 2\pi^2$ corresponding to 
contour lengths much larger than the contact radius, $L\gg R_{co}$, 
shape ${\rm I}$ can be approximated by  two semicircles of 
contact radius $R_{co}$ connected by two straight segments,
one of which is adhered to one step edge. This results in 
$\bar{E}_{\rm I} \approx -|w_{st}|\bar{L}/2 +3\pi|w_{st}|^{1/2}/\sqrt{2}$.
for the total energy.
The exact calculation in the appendix gives  a total energy
\begin{equation}
  \bar{E}_{\rm I} \approx -|w_{st}|\bar{L}/2 +4\sqrt{2}|w_{st}|^{1/2},
\label{EtotI}
\end{equation} 
for $|w_{st}|\bar{L}^{2} \gg 2\pi^2$,
which only differs in one prefactor.

Shape ${\rm II_{0}}$  can be  approximated by two 
  semicircular caps of diameter $a_{st}$, which  contribute a bending energy 
 $\bar{E}_b \approx 2\pi$, connected by  two straight adhered segments 
of length $L_{st} = L -\pi a_{st}$. This gives a total energy
\begin{equation}
  \bar{E}_{\rm II_0} \approx 2\pi -|w_{st}|(\bar{L} - \pi)
\label{EtotII0a}
\end{equation}
  for the total 
energy of shape ${\rm II_{0}}$.
Note that shape   ${\rm II_{0}}$ can only be realized for rings with 
$\bar{L}>\pi$ such that the ring can touch both step edges if it assumes 
a circular form. 
For intermediate adhesion strengths $|w_{st}|\approx 2$, this result 
agrees with the exact calculation. 
For  weak adhesion strengths $|w_{st}|\ll 1$, 
the exact calculation in the appendix gives a total energy
\begin{equation}
  \bar{E}_{\rm II_0} \approx 5.74 -|w_{st}| (\bar{L}-4.38)
-1.31|w_{st}|^{3/2},
\label{EtotII0}
\end{equation}
which differs only slightly in the numerical prefactors.

The energy branches corresponding to shapes ${\rm I}$ and ${\rm II_0}$ 
cross resulting in a {\em discontinuous} morphological 
transition between these two shapes  with 
 a jump in the adhered length $L_{st}$.
In the vicinity of this transition also 
the shapes ${\rm II_1}$ and ${\rm II_2}$  become stable or  metastable, 
which develop from the elongated shape 
${\rm II_0}$ by the formation of one and two {\em bulges},
 respectively. 
We can 
approximate the bulge by a semicircle with diameter $d_{\rm bul}>a_{st}$ 
and a weakly bent desorbed segments connecting this semicircle to the 
stripe edge, and determine the diameter $d_{\rm bul}$ and the length of the
 desorbed segment by optimizing the sum of
 bending energy gain and adhesion energy cost. 
For small $|w_{st}|$, this approximation gives an 
optimal bulge diameter  $d_{\rm bul}\sim a_{st}|w_{st}|^{-1/2}\sim R_{co}$,
which 
is proportional to the contact radius,
and  a total energy cost 
$\Delta \bar{E}_{\rm bul} \approx 
-\pi + 7.46\, |w_{st}|^{1/2} -2.92\,|w_{st}|$ for creating one bulge 
starting from the confined shape ${\rm II_0}$, 
which  is independent of $L$.

The exact calculation in the appendix gives 
\begin{equation}
  \Delta \bar{E}_{\rm bul} 
    \approx   -2.87+  7.50 |w_{st}|^{1/2}  -  3.72 |w_{st}|
 +0.30|w_{st}|^{3/2} 
  \label{DeltaEbul}
\end{equation}
for $|w_{st}|\ll 1$, which only slightly differs in the 
numerical prefactors. 
The resulting energies 
of both  bulged shapes  ${\rm II_m}$ are exactly 
given by  $\bar{E}_{\rm II_m} = \bar{E}_{\rm II_0} +
 m \Delta \bar{E}_{\rm bul}$ ($m=0,1,2$) as shown in the appendix. 
The bulges are thus localized and non-interacting excitations 
of shape ${\rm II_0}$. The energy branches of shapes ${\rm II_m}$ 
thus all intersect in a  single point in the bifurcation diagram 
Fig.~\ref{eminv}(a,b), which is determined by the condition 
$\Delta \bar{E}_{\rm bul}=0$. 
For $\Delta \bar{E}_{\rm bul}<0$, shape ${\rm II_2}$ with two bulges 
is energetically favorable, whereas for  $\Delta \bar{E}_{\rm bul}>0$,
shape ${\rm II_0}$ without bulges is energetically favorable. 
Therefore, shape ${\rm II_1}$ with one bulge is {\em never} globally stable
and, thus, there are only bulging transitions from 
shape  ${\rm II_0}$ directly  into  shape ${\rm II_2}$. 
Moreover, the shape transitions from shape ${\rm II_2}$ into 
shapes ${\rm II_0}$ or ${\rm I}$ are {\em discontinuous}.

We also show in the appendix analytically
 that bulges are only metastable for 
a limited range of adhesion strengths 
$w_{\rm min}(\bar{L}) <|w_{st}|< w_{\rm max}$.
For the upper stability limit we find 
the universal value  $w_{\rm max}\approx  0.35$, 
which holds both for shape ${\rm II_1}$ with a single bulge 
and shape ${\rm II_2}$ with two bulges. 
For $|w_{st}|> w_{\rm max}$, 
bulges become unstable with respect shrinking to zero size, 
 and 
the bulged shapes become unstable with respect to  
a spontaneous transition into shape ${\rm II_0}$. 
The lower stability limit $w_{\rm min}(\bar{L})$ depends 
on the contour length of the ring and slightly differs for shapes 
${\rm II_1}$ and ${\rm II_2}$. For $|w_{st}|<w_{\rm min}(\bar{L})$,
the bulges become so large that the adhered length 
on one or both of the surface steps shrinks to zero length, and 
the bulged shapes become unstable with respect to  
a spontaneous transition into shape ${\rm I}$.
  For $\bar{L}=20$ as in 
the bifurcation diagram  Fig.\ \ref{eminv}(a,b), we find 
$w_{\rm min} \approx 0.10$ for shape ${\rm II_2}$ and
 $w_{\rm min} \approx 0.07$ for shape ${\rm II_1}$.

The energies of the four metastable shapes can be probed in 
an ensemble  of adsorbed polymer rings of equal length $L$. 
The relative frequency of each shape 
is proportional to the Boltzmann weight associated
with its  energy.

\begin{figure}
\begin{center}
\includegraphics[width=0.46\textwidth]{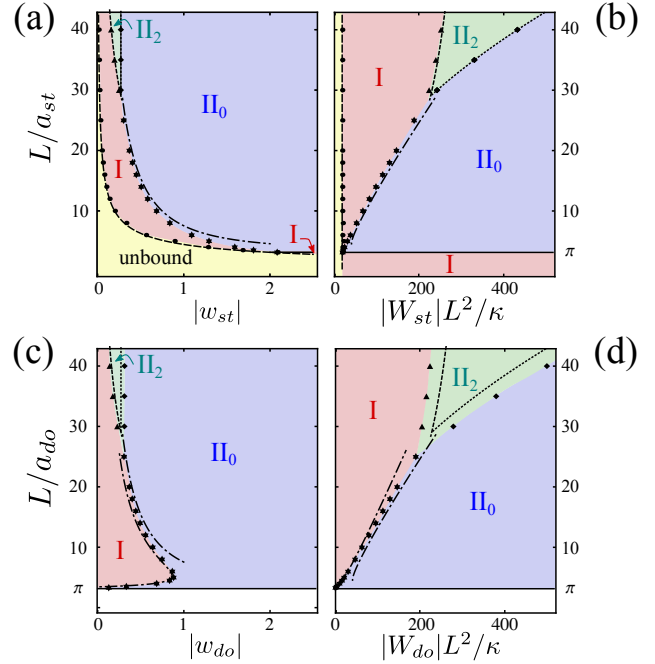}
\caption{\label{pd}  
Morphology diagram for a ring of length $L$ adhering to 
a {\em topographical surface groove} of width $a_{st}$ as a function of
(a) the contour length $L/a_{st}$ and the reduced potential 
strength $|w_{st}|$ as defined in (\ref{wst}) 
and (b) $L/a_{st}$ and $|W_{st}| L^2 /\kappa= |w_{st}|L^2/a_{st}^2$.
For  a {\em chemical 
domain} of width $a_{do}$, the morphological 
diagram is shown as a function of  (c) $L/a_{do}$
and  $|w_{do}|$ as defined in (\ref{wdo}) or (d) $L/a_{do}$
and  $|W_{do}| L^2 / \kappa$.
The parameter choice in (b) and (d) is advantageous 
if the structure width $a_{st}$ and $a_{do}$ is varied,
while the other system parameters are kept constant.
Morphological transitions as obtained from analytical 
energy minimization in (a) and (b)
and from numerical energy minimization in (c) and (d)
are represented by stars, triangles, diamonds and dots.
The  results (\ref{I_II0}), (\ref{bulge}), 
and (\ref{I_II2}) for these transitions are
indicated as dot-dashed, dotted and dashed lines, respectively. 
In (a) and (b) the dashed
line marks the unbinding transition,
whereas in (c) and (d) the dashed line marks 
the reentrant transition for small rings.
}
\end{center}
\end{figure}

\subsection{Morphology diagram}

The full morphology diagrams  Fig.\ \ref{pd}(a,b) shows how the 
stability of the four shapes
 ${\rm I}$, ${\rm II_0}$, ${\rm II_1}$ and ${\rm II_2}$ is controlled
by the parameters  $|W_{st}|$,  $L$, and $a_{st}$. 
Phase boundaries from numerical minimization are denoted by symbols.
The dashed line  corresponds to the continuous
unbinding transition (\ref{unbinding}) and  agrees 
with the numerical results.
The main feature of the morphology diagrams  is the
discontinuous transition
between morphologies ${\rm I}$ and ${\rm II_0}$,
(stars). Going through this transition with  increasing $|w_{st}|$,
 the ring goes from the 
round toroidal configuration ${\rm I}$ into the confined 
elongated configuration  ${\rm II_0}$.
The location of this transition can 
be derived from the condition $\bar{E}_{\rm I} =\bar{E}_{\rm II_0}$.
Using the estimates (\ref{EtotI}) and (\ref{EtotII0a}) 
    given above we find a  transition line 
\begin{equation}
  \bar{L}_{\rm I-II_0} \approx  
 2\pi  -  8 \sqrt{2}|w_{st}|^{-1/2}  + 4\pi |w_{st}|^{-1}
\label{I_II0}
\end{equation} 
valid for $|w_{st}| \lesssim 2$ and shown as dash-dotted line 
 in Fig.\ \ref{pd}(a,b).

This transition line terminates at $\bar{L}=\pi$ and $|w_{st}|=2$  where 
it intersects the unbinding transition line. 
For strong adhesion with 
$|w_{st}|>2$ a short ring first adheres  to one corner of the groove 
for $\bar{L} = \bar{L}_{ub}<\pi$ in a round shape ${\rm I}$, which, 
however,  lies completely {\em inside} the groove.
Only for  $\bar{L} > \pi$, the ring  touches the opposite corner   
of the groove and immediately assumes the elongated 
shape ${\rm II_0}$ because shape ${\rm I}$ is unstable in this 
regime,   see Fig.\ \ref{pd}(a,b).

At  the  transition 
line between configurations  ${\rm II_0}$ and 
${\rm II_2}$ (diamonds), which occurs for large $\bar{L}$,  
it becomes energetically favorable to form bulges on top 
of the confined shape ${\rm II_0}$, i.e.,  the energy
difference $\Delta \bar{E}_{\rm bul}$ becomes negative.
Also this bulging transition is discontinuous. 
Because both the caps of shape ${\rm II_0}$ and the 
bulges of shape ${\rm II_2}$ have a finite length, the 
bulge energy $\Delta \bar{E}_{\rm bul}$ only depends on $|w_{st}|$ and 
is independent of the contour length $L$. Therefore 
the transition line between between configurations  ${\rm II_0}$ and 
${\rm II_2}$ is also  independent of $L$ and, thus, 
{\rm vertical} in Fig.\ \ref{pd}(a). 
The exact location of the bulging transition  line can be found 
numerically by equating the exact  parametric representations of the 
total energies $\bar{E}_{\rm tot,II_0}$ and 
$\bar{E}_{\rm tot,II_2}$   and  the corresponding parametric 
representations  of $|w_{st}|$ 
given in the appendix, which gives
\begin{equation}
   |w_{\rm II_0-II_2}| \simeq 0.27.
\label{bulge}
\end{equation} 
and is shown as  dotted line in Fig.\ \ref{pd}(a,b).
Stable bulges form 
if the energy of a single bulge is negative 
for $|w_{st}|<|w_{\rm II_0-II_2}|$.
Because
 bulges are non-interacting, it is always energetically
 favorable to create two bulges  such that
shape  ${\rm II_1}$ is only metastable and 
absent in the morphology  diagrams, as already mentioned above.

The bulged configuration  ${\rm II_2}$ 
can become stable only above a {\em triple point}, where 
the three shapes ${\rm I}$, ${\rm II_0}$, and ${\rm II_2}$ coexist.
The  exact location of the triple point can be found numerically 
by equating the parametric representations of the 
total energies $\bar{E}_{\rm tot,I}$, $\bar{E}_{\rm tot,II_0}$, and 
$\bar{E}_{\rm tot,II_2}$ and the corresponding parametric 
representations  of $|w_{st}|$ 
given in the appendix. 
This gives  
\begin{equation}
  |w_{\rm tri}|=|w_{\rm II_0-II_2}| ~~\mbox{and}~~
\bar{L}_{\rm tri} \simeq   29.2.
\label{trip}
\end{equation}
Only sufficiently large rings $\bar{L}>\bar{L}_{\rm tri}$ 
can undergo a bulging transition. For shorter rings bulges are 
unstable and the ring directly assumes  the round shape 
${\rm I}$. 

For short rings $\bar{L}>\bar{L}_{\rm tri}$, 
we can obtain the transition line between shapes ${\rm II_2}$
and ${\rm I}$ (triangles) using the condition  $\bar{E}_{\rm I} 
=\bar{E}_{\rm II_2}=\bar{E}_{\rm II_0}+2\bar{E}_{\rm bul}$.
Using the estimates (\ref{EtotI}),  (\ref{EtotII0})  and (\ref{DeltaEbul})
    given above we find a  transition line
\begin{equation}
  \bar{L}_{\rm I-II_2} \approx 
      -6.13+ 18.68 |w_{st}|^{-1/2}   -1.44 |w_{st}|^{1/2}
\label{I_II2}
\end{equation} 
for $|w_{st}| \ll 1$
shown as dashed line in Fig.\ \ref{pd}(a,b).

The results (\ref{bulge}) and (\ref{trip}) for the bulging transition
and the triple point  are exact and 
the approximate formulas (\ref{I_II0}) for the transition 
between shapes ${\rm I}$ and  ${\rm II_0}$ and (\ref{I_II2}) 
for the  transition 
between shapes ${\rm I}$ and ${\rm II_2}$
are in good agreement with the numerical results and exact analytical 
results derived in  the appendix as can be seen 
in  the morphology diagrams in   Fig.\ \ref{pd}(a,b). 

The transition lines depend on the control parameters of the system, hence,
measurements of these transition lines can be used to determine material
parameters, such as $|W_{st}|$ or $\kappa$, experimentally. 
 As opposed  to other experimental methods to determine the bending
rigidity  no {\em external} forces, e.g.\ via an AFM tip, have to be
applied to the polymer, but the substrate pattern itself exerts forces on
the ring. 
In the morphology diagram Fig.\ \ref{pd}(a) we use the reduced 
contour length $\bar{L}$ and the reduced adhesion strength 
$|w_{st}|$ as control parameters. 
 In an experiment, the transition lines in the morphological diagram
Fig.\ \ref{pd}(a) are crossed in horizontal direction by
changing the adhesion strength $|W_{st}|$ of the substrate, which 
could be achieved by changing the substrate chemistry or surface 
charge. 
On the other hand, one could use rings of different
length on the same substrate and thereby observe ring morphologies along a
vertical line in Fig. \ref{pd}(a).  The last and maybe simplest
experiment is  to fabricate substrates with several grooves of different
width $a_{st}$. In this case, one would change the ratio of the involved
length scales $\bar{L}$ and the reduced adhesion strength $|w_{st}|$ at the
same time. Therefore, it is much more convenient to characterize the
system by $\bar{L}$ and the control parameter $|W_{st}| L^2/\kappa =
|w_{st}|\bar{L}^2$ as in the morphology diagram Fig.\ \ref{pd}(b). 
Changing the groove width and thus $\bar{L}= L/a_{st}$ 
corresponds to a vertical trajectory in this diagram.

All shape transitions between shapes ${\rm I}$,  ${\rm II_0}$, and 
${\rm II_2}$ are discontinuous. Therefore shapes remain metastable 
over a considerable parameter range, which gives rise to 
strong shape hysteresis along if any transition line is crossed 
in the morphology diagrams Fig.\ \ref{pd}(a,b). 
For a shape transition from a metastable state to another metastable
or stable  state 
a transition state corresponding to a saddle 
in the energy landscape has to be crossed. 
For some of the transitions 
 this transition state should also belong to one of the four 
classes of shapes ${\rm I}$, ${\rm II_0}$, ${\rm II_1}$, or 
 ${\rm II_2}$. For example, state ${\rm II_0}$ remains metastable 
down to adhesion strengths $|w_{st}|=0$. 
The transition states for transitions from ${\rm II_0}$ into 
 states ${\rm II_i}$ should also be bulged states.
Starting from shape ${\rm II_0}$,  
bulged states  form by crossing the 
maxima ${\rm II_1^*}$ or ${\rm II_2^*}$ containing 
{\em small} unstable bulges.  
The shape  ${\rm II_2^*}$
corresponds to a shape with two identical small
 bulges which are  unstable with respect to shrinking 
to zero size to a shape ${\rm II_0}$ or to expanding to its 
equilibrium size in state ${\rm II_2}$. 
Likewise, the shape  ${\rm II_1^*}$ contains a single small bulge which 
is unstable with respect  shrinking 
to zero size to a shape ${\rm II_0}$ or to expanding to its 
equilibrium size in state ${\rm II_1}$.
There will be an additional transition state between states 
${\rm II_1}$  to ${\rm II_2}$
containing  one small and one large 
bulge, where the small bulge is unstable with respect to shrinking 
to zero size to shape ${\rm II_1}$ or expanding to equilibrium size 
to state ${\rm II_2}$.

State ${\rm I}$ remains metastable for large adhesion 
strengths until the  round unbound segment
touches  the opposite corner of the groove in 
its midpoint,
which happens for $|w_{st}| \simeq 14.2$. 
For  transitions between states  ${\rm I}$ and  ${\rm II_i}$, 
where the round unbound segment attaches to the opposite 
corner of the groove, the transition state will 
presumably not fall into one of the four classes of stable states. 
For large $|w_{st}|$ state  ${\rm I}$ will attach  to the second 
corner of the groove by deforming asymmetrically.

\subsection{Thermal fluctuations}

The transition states represent local maxima in the energy landscape. 
Energy differences between the transition state and the corresponding 
minima give energy  barriers for shape transformations.
Thermal fluctuations allow the polymer ring to overcome these
energy barriers if 
$\Delta E < T$ ($k_B\equiv1$), which is equivalent to 
 $\bar{\Delta E} < 2a_{st}/L_p$ for a semiflexible 
 polymer with a persistence length $L_p = 2\kappa/T$.
Therefore, the influence of thermal fluctuations crucially depends 
on the ratio $L_p/a_{st}$: Our results apply 
for persistence lengths much larger 
than the stripe width, $L_p/a_{st}\gg 1$, where energy barriers
are relevant. Then all four (meta-)stable ring shapes are observable, 
and their morphological transitions exhibit  
a pronounced hysteretic behaviour.
For flexible polymers with $L_p/a_{st} \ll 1$, on the other hand, 
thermal fluctuations allow the polymer to change orientation within
the groove such that the four morphologies can no longer be clearly
distinguished.

A single surface step represents a potential well 
of depth $|W_{st}|$ and width 
$\ell$ comparable to the polymer diameter. Strong thermal fluctuations 
can give rise to a thermal 
unbinding transition from a single step if the potential 
strength  is smaller than a critical value,
$|W_{st}| < |W_{st,c}| = c  T/L_p^{1/3}\ell^{2/3}$ with a prefactor $c$ of
order unity \cite{KL03,kierfeld06}, which is equivalent to 
a critical value
\begin{equation}
  |w_{st}|<|w_{st,c}| = 2c a_{st}^2/L_p^{4/3}\ell^{2/3}.
\label{thunbinding}
\end{equation}
 for the reduced adhesion strength.
For $|W_{st}| < |W_{st,c}|$  binding to  surface steps
is prevented by thermal fluctuations and no morphological 
transitions can be observed. 
Before but close to thermal 
unbinding,  $|W_{st}| \lesssim |W_{st,c}|$, a reduced
{\em free} binding energy per length 
$f_{st} \sim W_{st} - W_{st,c}$, which 
includes also entropic contribution should  be used instead of the 
bare adhesion strength  $W_{st}$ \cite{kierfeld06}. 

All  morphological transitions derived above in the absence of thermal
fluctuations happen for 
reduced adhesion strength of order unity. Therefore, 
these results apply also in the presence of thermal fluctuations 
under the condition  $|w_{st,c}| \ll 1$, such that 
 thermal unbinding 
does not interfere with  the morphological transitions. 
The condition  $|w_{st,c}| \ll 1$ is equivalent to sufficiently 
small 
polymer diameters $\ell/a_{st}  \ll a_{st}^2/L_p^2$
or sufficiently wide grooves $a_{st} \gg \ell^{1/3}L_p^{2/3}$.
For typical polymer diameters in the nanometer regime and persistence 
lengths in the range of 50nm (DNA) up to 10$\mu$m (filamentous actin), 
this condition is fulfilled for  groove  widths $a_{st}\gg 10$nm for DNA
 and $a_{st}\gg 500$nm   for filamentous actin, respectively.
Even if morphological transitions are not modified, 
thermal unbinding can  preempt
 the bending energy induced 
 ring unbinding transition with $\bar{L}_{ub} \propto |w_{st}|^{-1/2}$,
see  eq.\ (\ref{unbinding}), but 
 only for large ring contour lengths 
$\bar{L} > |w_{st,c}|^{-1/2}$.

\section{Chemically structured striped
  surface domain}

The chemically structured stripe of width $a_{do}$ is modeled by
an additional adhesion energy gain $W_{do}<0$ per polymer
length for $|x|\le a_{do}/2$, which leads to a 
 generic square well adsorption potential with
 $V_{do}(x) = 0$ for $|x|> a_{do}/2$ and  
$V(x) = W_{do}$ for $|x| \le a_{do}/2$.
The adhered length $L_{do}$ is given by the polymer length
within the stripe  $|x|\le a_{do}/2$, and the adhesion energy is 
$E_{ad} = -|W_{do}|L_{do}$. 

The analytical energy minimization becomes involved for the 
chemical surface domain because adhered segments are no 
longer perfectly straight as for the topographical groove. 
Therefore, we 
performed the  energy minimization numerically 
using the SURFACE EVOLVER. 
We find the same  four types of morphologies ${\rm I}$, ${\rm II_0}$,
${\rm II_1}$, and ${\rm II_2}$  as for the topographical 
groove, see Fig.~\ref{types}(b).
Also for the chemically structured stripe,  there are two 
possible shapes ${\rm  II_2}$, one with bulges on the same side 
and one with bulges on opposite sides. Both configurations 
have very similar energies but are no longer 
strictly degenerate as for the topographical groove:
The antisymmetric 
shape with bulges on opposite sides as shown in 
Fig.~\ref{types}(b) has a slightly lower energy 
for the chemically structured stripe. 

Remarkably, 
 ring shapes minimizing the bending energy are almost 
identical as compared to   a topographical groove 
of the same width $a_{st}=a_{do}$, see Fig.\ \ref{types}. 
Furthermore, the bending energies of constrained  equilibrium 
shapes  agree to a good approximation if the adhered length $L_{do}
\approx L_{st} + \Delta L_{ad}$ is shifted  by a   constant amount 
 $\Delta L_{ad}$:
 In contrast to the  groove, the stripe domain 
is also adhesive  between its boundaries for 
$|x|< a_{do}/2$ such that  the same ring shape  has  a larger  
adhered length. 
As a result of this shift, 
the bifurcation diagram for the total  energies of all local
extrema as a function of the reduced adhesion strength 
\begin{equation}
  |w_{do}| \equiv
|W_{do}|a_{do}^2/\kappa.
\label{wdo}
\end{equation} 
resembles
 the corresponding diagram Fig.~\ref{eminv} for a 
surface groove. 

For shape ${\rm II_0}$ we can show this quantitatively 
because the total energy of shape ${\rm II_0}$ can be exactly 
calculated for the chemical domain, 
\begin{equation}
   \bar{E}_{\rm II_0} =  5.74 -|w_{do}|\bar{L} 
\label{EtotII0do}
\end{equation}
where we used the reduced length 
and reduced energies
\begin{equation}
   \bar{L}\equiv L/a_{do} ~~\mbox{and}~~\bar{E} \equiv Ea_{do}/\kappa.
\label{barE2}
\end{equation}
analogously 
to the topographical groove. 
As compared to the the result (\ref{EtotII0}) for 
the topographical groove we notice the 
agreement in the limit of small small adhesion strengths 
with a constant  shift $\Delta \bar{L}_{ad} = 4.38$ of the 
adhered length. For the shape ${\rm II_0}$ this shift corresponds 
exactly to the length of the curved caps in the limit of weak 
adhesion as discussed in the appendix.

As a result of this approximate 
mapping of shapes and energies between the two 
types of adhesive stripes,  the   morphology diagrams in the plane 
spanned by the reduced potential strength $|w_{do}|$ and 
 $\bar{L}$, as shown in  Fig.\ \ref{pd}(c,d),
look very similar for the chemical stripe  
and the  topographical surface groove. 
In particular, our results (\ref{I_II0}) 
for the  transition between 
shapes ${\rm I}$ and ${\rm II_0}$ (dash-dotted line in Fig.\  \ref{pd}(c,d)), 
(\ref{bulge}) for the appearance of bulged states (dotted line), 
and (\ref{I_II2})  for the transition between shapes  
${\rm I}$ and ${\rm II_2}$ (dashed line)
remain valid and  agree well 
with the numerical results 
(stars, diamonds, and triangles, respectively).
 
However, the unbinding transition of shape ${\rm I}$ 
is absent for the chemical stripe domain: 
It is always energetically favorable 
for the ring to adhere to the striped domain. 
Furthermore, the two phase diagrams
 differ in the behavior of small rings.
Small rings can fully bind  to the
chemical stripe without  deformation and shapes ${\rm I}$ and ${\rm
  II_0}$  become equivalent, which 
 leads to the  re-entrance of shape ${\rm II_0}$ close to 
 $\bar{L}=\pi$. 
We estimate the location of this re-entrant transition  can
by approximating  small rings in  shape ${\rm I}$ 
by a circle. The adhered length of such a circle is 
$\bar{L}_{do} = \bar{L}\arccos(1-2\pi/\bar{L})/\pi $
 so that the total  energy is  
\begin{equation}
\bar{E}_{\rm I} \simeq 2 \pi^2/\bar{L}  - |w_{do}|\bar{L}_{do}.
\label{EtotIdo}
\end{equation}
Equating this energy with the total energy (\ref{EtotII0do})
of shape ${\rm II_0}$ we find 
\begin{equation} \label{I_II0_small}
|w_{\rm I-II_0}| \simeq 
  \frac{\pi 
 (5.74 \bar{L}-2\pi^2)}{ \bar{L}^{2}
\arccos \left(2 \pi/\bar{L} -1\right)}
\end{equation}
for $\bar{L}>3.44$. This result is also 
shown as dash-dotted line in Fig.\  \ref{pd}(c,d)
and gives remarkable agreement with the exact numerical results (stars) 
also for lengths up to the triple point. 
Because of the re-entrance  the elongated shape 
${\rm II_0}$ is the stable state for 
adhesion strengths $|w_{do}|>0.86$.

\section{Periodic stripe structures}

\begin{figure}
 \begin{center}
 \includegraphics[width=0.46\textwidth]{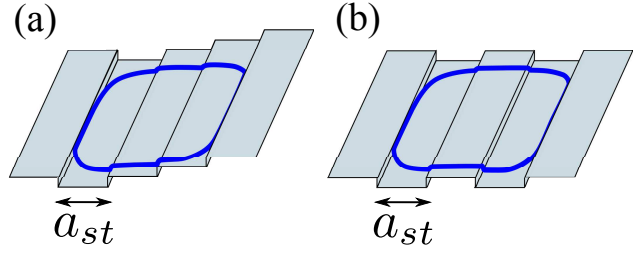}
\caption{\label{terraces}  
 Adsorbed polymer ring on two periodic stripe patterns (a) and (b) 
 consisting of 
several equidistant topographical surface steps with distance $a_{st}$. 
On both patterns (a) and (b),
the ring is shown in configuration  ${\rm II^3_0}$ connecting 
two surface steps at distance $3a_{st}$.
For small  surface step heights both configurations are
approximately identical. 
 }
 \end{center}
 \end{figure}

An important generalization of our system, which can serve as a
model for the atomic lattice structure of substrates, is 
 a {\em periodic} stripe pattern. 
Specifically, we consider an array of  equidistant 
parallel topographical 
surface steps located 
at $x= ia_{st}$  as they occur, e.g.\, on vicinal 
surfaces, see  Fig.\ \ref{terraces}.
For  surface step heights smaller or comparable 
to the polymer diameter we can  
neglect small energy corrections arising if the polymer
 crosses the surface steps. Then upward and downward steps
 have the same effect on ring shapes and the two surface step patterns
shown in Fig.\ \ref{terraces} give rise to 
approximately  identical  metastable 
ring morphologies with almost identical energies. 

Such surface structures drastically increase the number of metastable polymer
shapes. Before presenting  general results  for the full
 periodic pattern, we start by adding  a single parallel surface step to the
groove shown in Fig.~\ref{types}(a) at distance $a_{st}$. 
 The resulting metastable ring morphologies can be
classified into conformations that adhere to one (${\rm I}$), two (${\rm II}$)
or three (${\rm III}$) edges plus the unbound circular shape. 
 Clearly, the
shapes ${\rm I}$ are the same as for the single stripe, as the remaining steps
(up to small corrections) do not contribute to the energy and also the
unbinding transition applies without modifications 
to the three-step-geometry.

Moreover, if the ring binds to two edges, it should attain shapes that
correspond to the morphologies ${\rm II_0}$ and ${\rm II_2}$ we found
before but now
the ring can adhere either to two neighboring steps 
(at a distance $a_{st}$) or to the two outer
steps (at a distance $2a_{st}$).
 Formally, we will distinguish these two cases via a superscript
that indicates the distance between the
 relevant edges in units of $a_{st}$, 
i.e.,  ${\rm II^1_0}$, ${\rm II^2_0}$ etc. 
By analyzing the corresponding energy
estimates $E_{\rm tot}(|w_{st}|)$ one finds, that shape
${\rm II_0^2}$ is always energetically
 favorable compared to shape ${\rm II_2^1}$. 
Furthermore, shape ${\rm II_2^2}$ becomes only stable 
for very large rings, i.e. $L/2 a_{st} > \bar{L}_{\rm tri}\simeq 29.2$.
Therefore, shapes with bulges  
can be neglected altogether to a good approximation.

Now we address the full periodic stripe pattern. 
Neglecting the formation of bulges
the possible stable states  are  shape
${\rm I}$ and elongated shapes ${\rm II_0^n}$ where
the ring binds to two surface steps $i$ and $i+n$ 
(shape ${\rm II_0^1}$ is 
  identical to shape ${\rm II_0}$). All shapes ${\rm II_0^n}$
 can be approximated by two
  semicircles of diameter $na_{st}$, 
connected by  two straight adhered segments 
of length $L_{st} = L -n\pi a_{st}$, which gives a  total 
energy 
\begin{equation}
  \bar{E}_{\rm II_0^n} \approx 2\pi/n -|w_{st}|(\bar{L} - n \pi).
\end{equation} 
For a ring of contour length $L$, only states with $n\le
n_{\rm max}=[\bar{L}/\pi]$ are accessible.
At small $|w_{st}|$ the round shape ${\rm I}$ is stable. 
The criterion $\bar{E}_{\rm I} =\bar{E}_{\rm II_0^{n_{\rm max}}}$  
gives a first  transition at 
\begin {equation}
|w_{\rm I-II_0^{n_{\rm max}}}|\approx \frac{2\pi}{\bar{L}n_{\rm max}}
\label{wIIIn}
\end{equation}
 from
shape I into shape  ${\rm II_0^{n_{\rm max}}}$.
The criterion $\bar{E}_{\rm II_0^{n+1}}=\bar{E}_{\rm II_0^n}$
 gives a cascade 
of $n_{max}-1$ further  morphological transitions from shape 
 ${\rm II_0^{n+1}}$ into shape  ${\rm II_0^{n\vphantom{+1}}}$ 
 at adhesion strengths 
\begin{equation}
  |w_{\rm II_0^n-II_0^{n-1}}| \approx 
\frac{2}{n(n-1)}  ~~~(2\le n \le  n_{\rm max}),
\label{wnn-1}
\end{equation}
which are  {\em independent} of the contour length $L$. 
For strong adhesion 
$|w_{st}|>|w_{\rm II_0^2-II_0^{1}}|=1$,
 shape  ${\rm II_0^{1}}$ remains the ground state.

In the limit $|w_{st}|\gg 1$, 
the contact radius becomes small compared to the 
distance $a_{st}$ between surface steps such that 
shapes ${\rm II_0^{n}}$ become effectively kinked 
with sharp bends. 
According to the result (\ref{wnn-1}), only shape 
 ${\rm II_0^{1}}$ connecting neighboring surface steps 
is stable in this limit and, thus,  sharp kinks 
should be observable only for this shape.

\begin{figure}
\begin{center}
\includegraphics[width=0.46\textwidth]{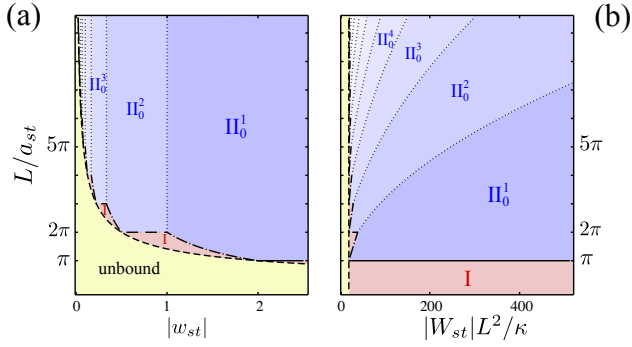}
\caption{\label{fig:morph_diag_nsteps} 
  Morphology diagrams of a ring adhering to a substrate
  with  $n$  equidistant surface topographical 
  steps at  distance $a_{st}$
  as a function of (a) the contour length $L/a_{st}$ and the reduced potential 
strength $|w_{st}|$ as defined in (\ref{wst}) 
and (b) $L/a_{st}$ and $|W_{st}| L^2 /\kappa= |w_{st}|L^2/a_{st}^2$.
  Shapes  ${\rm II_0^n}$
   with $n\ge2$ 
represent elongated rings which bind to two surface steps that 
are separated by $n$ terraces: the stability regimes of these shapes 
are  shown in 
  different colors of blue.  The dashed line marks the unbinding transition,
    and the dot-dashed  and dotted lines indicate the
  morphological transitions as estimated in eqs.\ 
     (\ref{wIIIn}) and (\ref{wnn-1}),
  respectively.  }
\end{center}
\end{figure}

We summarize our findings in a morphology diagram in Fig.\ 
\ref{fig:morph_diag_nsteps} using the 
control parameters $\bar{L}$ and $|w_{st}|$ in 
 Fig.\ \ref{fig:morph_diag_nsteps}(a) or  $\bar{L}$ and
 $|W_{st}| L^2/\kappa =
|w_{st}|\bar{L}^2$  in 
 Fig.\ \ref{fig:morph_diag_nsteps}(b).
Changing the adhesion strength $|w_{st}|$ corresponds 
to horizontal paths in  Figs.\ \ref{fig:morph_diag_nsteps}(a,b)
and results in a cascade of transitions between 
different shapes  ${\rm II_0^{n}}$. 
The same transition cascade is encountered when 
changing the surface step spacing and thus $\bar{L}= L/a_{st}$ 
corresponding to a 
vertical line in  Fig.\ \ref{fig:morph_diag_nsteps}(b).

For periodic stripe structures thermal fluctuations can play 
an important role. As discussed already for the surface groove 
they can give rise to a complete thermal unbinding from 
individual surface steps such that no morphological 
transitions can be observed below  the critical reduced  
adhesion strength $|w_{st,c}|$, see eq.\ (\ref{thunbinding}).
For a periodic stripe pattern, all morphological transitions  
$|w_{\rm II_0^n-II_0^{n-1}}|\sim 1/n^2$ with $n< |w_{st,c}|^{-1/2}$ 
should be observable before  thermal unbinding happens. 
As discussed above we find $|w_{st,c}|\ll 1$ for small 
polymer diameters and sufficiently large distances 
$a_{st} \gg \ell^{1/3}L_p^{2/3}$  between 
surface steps.

Another effect of thermal fluctuations are 
{\em additional}  kink-like excitations connecting 
neighboring surface steps
 \cite{kraikivski04,kraikivski05,hochrein06,hochrein07}.
In the absence of thermal fluctuations such kink excitations 
are absent as they cost an additional 
kink  energy $E_{\rm kink}$. Thermal fluctuations  create 
kinks with an average density 
\begin{equation}
  \rho_{\rm kink} \sim e^{-E_{\rm kink}/T} = 
    e^{-\bar{E}_{\rm kink}L_p/2a_{st}}
\label{rhokink}
\end{equation}
   along the ring contour 
\cite{kraikivski05}.

If step distances are  small compared to the contact radius,
$a_{st} \ll R_{co} = (\kappa/2|W_{st}|)^{1/2}$ or $|w_{st}|\ll 1$, 
the  kink is elongated with  a length  $L_{\rm kink} \sim
a_{st}^{1/2}R_{co}^{1/2} \sim a_{st}|w_{st}|^{-1/4}$   along the surface
steps. 
The kink energy is $E_{\rm kink} \sim a_{st}^{1/2}\kappa^{1/4}|W_{st}|^{3/4}$ 
or $\bar{E}_{\rm kink} \sim |w_{st}|^{3/4}$ in this regime
 \cite{kraikivski04,kraikivski05}. Therefore the thermal 
kink density is exponentially low according to eq.\ (\ref{rhokink}) 
if $|w_{st}| \gg (a_{st}/L_p)^{4/3}$, and kinks do not 
modify  morphological transitions 
 $|w_{\rm II_0^n-II_0^{n-1}}|\sim 1/n^2$ with 
$n< (L_p/a_{st})^{2/3}$. For persistence lengths much larger 
than the step distances, $L_p/a_{st}\gg 1$, a large cascade of 
transitions should remain observable. 
According to  eqs.\ 
     (\ref{wIIIn}) and (\ref{wnn-1}) most of the   
morphological transitions 
 for the periodic stripe pattern take place in the regime $|w_{st}| \ll 1$
corresponding to $a_{st} \ll R_{co}$.

If step distances are  large  compared to the contact radius,
$a_{st} \gg R_{co}$ or $|w_{st}|\gg 1$, 
the kink crosses the potential barrier of 
width  $a_{st}$ in a right angle with two small curved segments
of contact radius $R_{co}$ connecting to the surfaces edges.
\footnote{In Refs.\ \citenum{hochrein06,hochrein07} such kink 
excitations have been called ``crossings''.}
This gives rise to a kink length 
$L_{\rm kink} \sim a_{st}$ and a kink energy $E_{\rm kink} \sim a_{st}
|W_{st}|$
 or $\bar{E}_{\rm kink} \sim |w_{st}|\gg 1$.
Also in this regime the thermal 
kink density is exponentially low according to eq.\ (\ref{rhokink}) 
 for  persistence lengths much larger 
than the step distances, $L_p/a_{st}\gg 1$.

\section{Ring condensation}

Finally, our model is applied to the condensation transition 
of semiflexible polymer rings 
in poor solvent or in the presence of condensing agents giving rise 
 an  effective polymer-polymer attraction  with a 
 condensation energy gain $W_{\rm con}<0$  per contact length.

For small condensation energies, the semiflexible ring will remain 
in a round ring configuration with its total energy given by the 
bending energy 
\begin{equation}
E_{\rm ring} = E_{b, \rm ring} =  2\pi^2 \kappa/L.
  \label{Ering}
\end{equation}
For  strong attractive interactions between polymer segments, 
one  expects the polymer ring to form a toroid, 
similar to  open polymers \cite{bloomfield91,bloomfield97}.
The  radius of the toroid will be 
$L/2 \pi n$, where $n$ is the winding number and $L$ the contour length 
of the ring. 
In comparison to the ring adsorbed to the stripe structures, 
the length scale of the stripe width is absent, and the 
morphologies in the presence of a condensing potential are 
characterized by only {\em one} parameter, namely 
$|W_{\rm con}|L^2/\kappa$
(if $L_p$ is large such 
 that thermal unbinding can be  neglected \cite{KL03}).
The total energy of a toroidal configuration with 
winding number $n=2$  is 
\begin{equation}
    E_{\rm tor} = \frac{\kappa}{2L} \left( 16\pi^2 - 
       \frac{|W_{\rm con}|L^2}{\kappa}\right)
\label{Etor}
\end{equation}
For toroids with $n>2$ the packing structure, which is commonly assumed to 
be hexagonal \cite{bloomfield91,bloomfield97,schnurr2000,schnurr2002}, 
has to be taken into account.

Comparing the energies of toroids and rings, we find that 
a discontinuous  transition from a ring to the first condensed toroidal 
state with $n=2$ windings occurs at 
$|W_{\rm con,tor}|L^2/\kappa = 12 \pi^2$.

 \begin{figure}
 \begin{center}
\includegraphics[width=0.3\textwidth]{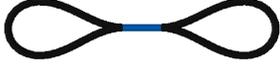}
 \caption{\label{racquet}
   Metastable racquet shape of a condensed ring.
 Segments that adhere to each other are colored in blue. 
}
 \end{center}
 \end{figure}

Finally the ring can also assume racquet-shaped metastable configurations
as shown in Fig.\ \ref{racquet},
which resemble the elongated shapes ${\rm II_2}$ containing two bulges. 
Performing a similar calculation as for the bulged shapes, 
which is contained in appendix, we 
can calculate the total energy of the  racquet shapes  as 
\begin{equation}
   E_{\rm tot,rac} = \frac{\kappa}{2L} \left(
     12.85 \left(\frac{|W_{\rm con}|L^2}{\kappa}\right)^{1/2} -
         \frac{1}{2} \frac{|W_{\rm con}|L^2}{\kappa}\right)
\label{Erac}
\end{equation}
which is valid for $|W_{\rm con}|L^2/\kappa> 73.33$ sufficiently large that 
the contact length is nonzero. 
Comparing with the energy (\ref{Etor}) of the toroidal configurations,
 we find that the racquet shape has  a higher energy for 
$|W_{\rm con}|L^2/\kappa > 82.49$, i.e., for all $|W_{\rm con}|>|W_{\rm con,tor}|$. 
Comparing with the energy  (\ref{Ering}) of the uncondensed ring,
 we find that the racquet shape has  a higher energy for 
$|W_{\rm con}|L^2/\kappa < 578.34$, i.e., for all $|W_{\rm con}|<|W_{\rm con,tor}|$.
Therefore, either the uncondensed ring or the toroidal configuration 
represent the global energy minimum, and 
racquet-like shapes  are only metastable configurations
in ring condensation. 
In contrast, the analogous bulged 
elongated shapes ${\rm II_2}$ can represent globally stable 
states of semiflexible rings adsorbed on striped substrates.

\section{Conclusion}

We showed that morphologies of  
adsorbed semiflexible polymer ring on a substrate containing an adhesive
stripe domain can be completely classified. Whereas a 
flexible polymer ring assumes a random coil configurations, which 
can easily adapt its shape to fit into the 
adhesive stripe as long as the  persistence length of the flexible polymer 
is much smaller than the 
stripe width, 
the bending energy of a semiflexible polymer ring leads to 
the existence of only four distinct metastable states as shown 
in Fig.\ \ref{types}:
A  round toroidal
 configuration ${\rm I}$, a confined elongated
shape ${\rm II}$, as well as two shapes ${\rm II_1}$ and ${\rm II_2}$ 
containing one or two bulges, respectively.

Specifically we considered two types of adhesive stripe 
domains,  
 topographical surface grooves  and  chemically structured 
   surface domains.
Both types of structures lead to very similar behavior:
a  {\em discontinuous morphological transition}
between the two dominant shapes ${\rm I}$ and ${\rm II_0}$, 
as well as  intermediate bulge shapes ${\rm II_1}$ and  ${\rm II_2}$ 
for large contour lengths, of which only shape ${\rm II_2}$ containing 
two bulges can be globally stable.   

Estimates for all transition lines 
were derived, see Fig.\ \ref{pd}, 
which could serve to determine material properties
of the substrate or the polymer ring experimentally.
The discontinuous transitions display shape hysteresis and 
are  observable  for persistence lengths exceeding
the stripe width.

For a periodic array of topographic steps we find a cascade of 
morphological shape transitions as displayed in Fig.\ 
\ref{fig:morph_diag_nsteps}.

\section{Acknowledgments}

We acknowledge financial support by the Deutsche Forschungsgemeinschaft 
via Sonderforschungsbereich  448.

\appendix

\section*{Appendix}

\section{Analytical energy minimization for topographical surface steps}
\label{appendix1}

In this appendix we derive exact analytical results for the 
metastable shapes of  a ring
 adhering to a topographical surface groove with two adhesive edges.
All four metastable ring shapes 
consist of one or two straight segments 
(with tangent angles $\theta(s) = 0$ or $\theta(s)=\pi$) 
of total length $L_{st}$, which  adhere to the 
straight stripe edges and one or two  curved segments of total 
length $L-L_{st}$.
The total energy of the ring is  
\begin{equation} \label{etot_general}
E_{\rm tot}(L_{st})=E_b(L_{st}) - |W_{st}| L_{st}.
\end{equation}
Only the curved segments  contribute to 
the bending energy 
$E_b = (\kappa/2)\int_0^{L} ds (\partial_s \theta(s))^2$, 
whereas only the straight adhered segment
contribute to  the adhesion energy $E_{ad}= -|W_{st}| L_{st}$.

For each shape 
 additional constraints have to be imposed for ring closure,
which take on slightly different forms for the shapes of type ${\rm I}$ 
adhering to one edge and shapes ${\rm II}$ adhering to both edges 
of the stripe. 

The total energy 
is minimized with respect to variations of the tangent angle 
configuration $\theta(s)$ with $0\le s\le L$.
For each metastable shape  we first minimize the bending energy 
under the additional constraint of 
fixed adhered length $L_{st}$ 
to obtain the constrained bending energy minimum 
 $E_b=E_b(L_{st})$ as a function of $L_{st}$. Then we 
minimize the total  energy (\ref{etot_general}) also with respect 
to $L_{st}$  to obtain the unconstrained minimal energy 
$E_{\rm tot}=E_{\rm tot}(W_{st})$ as a function of $|W_{st}|$, 
which  is equivalent to a Legendre transform of the bending 
energy $E_{b}(\bar{L}_{st})$ 
with respect to the adhered length $L_{st}$. 
We obtain exact results for energy minima $E_b=E_b(L_{st})$
and $E_{\rm tot}=E_{\rm tot}(W_{st})$  in parametric 
form and can solve for 
explicit formulae in the limits of strong and weak adhesion.

\subsection {Shape ${\rm I}$}

Shape ${\rm I}$  contains one round  segment of length $L_r$ and one
adhered segment with   $L_r+L_{st}= L$.
Shape ${\rm I}$ is parameterized with one half of the curved segment 
at arc lengths 
$0 \le s \le L_r/2$ as shown in Fig.\ \ref{fig:sketches}(a).
Considering one half of the symmetric configuration, 
the ring closure constraint
for the coordinate parallel to the groove can be written as 
\begin{equation}
   \int_{0}^{L_r/2}ds \cos\theta(s)  + \frac{L_{st}}{2} = 0.
\label{constraintI}
\end{equation}
We associate this constraint with a Lagrange multiplier $\mu$. 
The resulting Euler-Lagrange equation minimizing the bending 
energy of the round segment is 
\begin{equation}
   \kappa \partial_s^2\theta + \mu \sin\theta = 0
\end{equation}
Integrating once we find 
\begin{eqnarray}
 &&   \frac{\kappa}{2} (\partial_s\theta)^2 - \mu \cos\theta = c
\label{ELI}\\
 &&  ds = d\theta \left[ \frac{2\mu}{\kappa}( q+\cos\theta) \right]^{-1/2}
\nonumber
\end{eqnarray}
with an integration constant $c$ and 
 the parameter $q\equiv c/\mu$. We have  two  equations
for the two unknown parameters $q$ and $\mu$:
The first equation gives the length of the round segment
\begin{eqnarray}
   L_r &=& L-L_{st} = 2\left( \frac{2\kappa}{\mu} \right)^{1/2}f_1(q)  
\label{LrI}\\
   f_1(q) &\equiv & \int_0^\pi d\theta (q+\cos\theta)^{-1/2} 
\nonumber
\end{eqnarray}
The second equation is given by  the constraint (\ref{constraintI}) 
\begin{eqnarray}
   L_{st} &=& 2 \left( \frac{2\kappa}{\mu} \right)^{1/2}f_2(q)  
\label{LstI}\\
   f_2(q) &\equiv & - \int_0^\pi  d\theta\cos\theta (q+\cos\theta)^{-1/2} 
\nonumber
\end{eqnarray}
The functions $f_1(q)$ and $f_2(q)$ can be expressed by 
elliptic integrals, which converge for $q>1$.

Instead of solving explicitly for $q$ and $\mu$, 
we will express all quantities 
of interest parametrically as functions of $q$ using the two 
equations (\ref{LrI}) and (\ref{LstI}). 
This gives 
\begin{eqnarray}
  \frac{L_{st}}{L}  &=& 1- \frac{L_r}{L} = \frac{f_2(q)}{f_1(q)+f_2(q)}
\label{LstqI}\\
  \frac{2\mu L^2}{\kappa} &=& 4 (f_1(q)+f_2(q))^2
\label{muqI}
\end{eqnarray}
In addition, 
the bending energy can be rewritten as 
\begin{eqnarray}
   E_{b} &=&  \left( {2\kappa}{\mu} \right)^{1/2}f_3(q)  
\label{EbI}\\
   f_3(q) &\equiv &  \int_0^\pi d\theta (q+\cos\theta)^{1/2} 
\nonumber
\end{eqnarray}
where  also the function $f_3(q)$ can be expressed by elliptic integrals. 
It follows that  the bending energy is given by 
\begin{equation}
   \frac{E_b L}{\kappa}  =  2 (f_1(q)+f_2(q)) f_3(q).
\label{EbqI}
\end{equation}
Eqs.\ (\ref{EbqI}) and (\ref{LstqI}) give a parametric representation 
of $E_b(L_{st})$ using the parameter $q>1$. 
The corresponding curve is shown in Fig.\ \ref{fig:landscape}(a) as red line. 

For  $q\approx 1$, 
 both $f_1(q)$ and  $f_2(q)$ diverge while $f_1(q)-f_2(q) \approx f_3(q)
\approx 2\sqrt{2}$ , such that 
 $L_{st}\approx L/2$ and 
\begin{equation}
  \frac{E_b L}{\kappa}  \approx  8L (L/2-L_st)^{-1}
\end{equation}
diverges  corresponding to the 
limit of 
maximal adhered length and an maximally elongated ring configuration. 
In the limit of large $q \gg 1$, we find 
$f_1(q) \approx \pi q^{-1/2}$,  $f_2(q) \approx \pi q^{-3/2}$, and $f_3(q)
\approx  \pi q^{1/2}$ such that 
$L_{st} \approx 0 $ and $E_b L/\kappa  \approx  2\pi^2$ corresponding 
to a circular ring adhering in a single point.

If  the constraint of fixed adhered length $L_{st}$ is lifted, we have 
to minimize the total energy (\ref{etot_general})  also with respect to
variations of $L_{st}$. This  gives a transversality 
condition for the contact curvature at each contact point where 
a curved segment joins the straight adhered segments 
\begin{equation}
  |\partial_s\theta({s_{co}})|  = \frac{1}{R_{co}} = 
   \left( \frac{2|W_{st}|}{\kappa} \right)^{1/2}
\label{coI}
\end{equation}
Using this condition in (\ref{ELI}) we find $c= |W_{st}| -\mu$ or 
\begin{equation}
   q = \frac{|W_{st}|}{\mu} -1
\end{equation} 
which allows us to express also $|W_{st}|$ as a function of the parameter 
$q$ using (\ref{muqI}),
\begin{equation}
   \frac{|W_{st}|L^2}{\kappa}  = 2(q+1) (f_1(q)+f_2(q))^2,
\label{WqI}
\end{equation}
and to obtain together with (\ref{EbqI}) and (\ref{LstqI}) 
a parametric 
representation of $E_{\rm tot}(|W_{st}|)$ using the parameter $q$.
The corresponding curve is shown in Figs.\ \ref{eminv}(a,b) as red line.

The limiting case $q\approx 1$  with  $L_{st}\approx L/2$ 
corresponds to  adhesion strengths
$|W_{st}|L^2/\kappa = L^2/2R_{co}^2 \gg 2\pi^2$. Using the asymptotics 
of the functions $f_i(q)$ we find 
\begin{equation}
    \frac{E_{\rm tot,I}L}{\kappa} \approx   4 \sqrt{2} 
         \left(\frac{|W_{st}|L^2}{\kappa}\right)^{1/2}  - \frac{1}{2}
  \frac{|W_{st}|L^2}{\kappa} 
\label{Etotapp1I}
\end{equation}
in this limit. 
Large $q\gg 1$ with $L_{st} \approx 0 $ corresponds to 
 $|W_{st}|L^2/\kappa \approx 2\pi^2$, which is the critical 
value for the   unbinding transition from a single 
surface step, 
which is also known for  vesicles adhering to a
surface  \cite{seifertlipowsky1990,seifert1991}.
In the vicinity of this critical value we find 
\begin{equation}
    \frac{E_{\rm tot,I}L}{\kappa} \approx   2\pi^2 - 
     \frac{1}{24\pi^2}\left(\frac{|W_{st}|L^2}{\kappa}-2\pi^2  \right)^2
\label{Etotapp2I}
\end{equation}
which shows that the unbinding transition is {\em continuous}.
The asymptotic estimate  (\ref{Etotapp1I}) is
shown in Fig.\ \ref{eminv}(a) as dashed line.

\subsection{Shape ${\rm II_0}$}

Shape ${\rm II}_0$ consists of two  
round caps of lengths $L_{\rm cap,1}$ and 
$L_{\rm cap,2}$  and two adhered segments 
with total  length $L_{st}$ such 
that  $L_{\rm cap,1}+L_{\rm cap,2} + L_{st} =L$. 
The caps have reflection symmetry 
with respect to the axis  $x=a_{st}/2$.
The ring closure constraints for the coordinate 
perpendicular to the groove ensure that the curved segments 
connect both groove edges,
\begin{equation}
   \int_{L_{\rm cap,i}}ds\sin\theta(s)  -a_{st} = 0,
\label{constraintII0}
\end{equation}
for each cap $i=1,2$. 
The constraints are associated with Lagrange multipliers $\nu_i$. 
The Euler Lagrange equations for the shape of the caps 
become
\begin{equation}
   \kappa \partial_s^2\theta - \nu_i \cos\theta = 0
\label{EL1II0}
\end{equation}
Integrating once we find 
\begin{eqnarray}
 &&   \frac{\kappa}{2} (\partial_s\theta)^2 - \nu_i \sin\theta = c_i
\label{ELII0}\\
 &&  ds = d\theta \left[ \frac{2\nu_i}{\kappa}( p_i+\sin\theta) \right]^{-1/2}
\nonumber
\end{eqnarray}
with  integration constants $c_i$ and  parameters $p_i\equiv c_i/\nu_i$.
In total we have to determine 
 six  unknown parameters $L_{\rm cap,i}$, $p_i$, and $\nu_i$.
These parameters have to fulfill  four equations 
\begin{eqnarray}
   2L_{\rm cap,i} &=&  \left( \frac{2\kappa}{\nu_i} \right)^{1/2}
g_1(p_i)  
\label{LcapII0}\\
   g_1(p) &\equiv & \int_0^\pi d\theta (p+\sin\theta)^{-1/2} 
\nonumber
\end{eqnarray}
for the cap lengths and 
\begin{eqnarray}
   a_{st} &=& 2 \left( \frac{\kappa}{2\nu_i} \right)^{1/2}g_2(p_i)  
\label{astII0}\\
   g_2(p) &\equiv & - \int_0^\pi  d\theta\sin\theta (p+\sin\theta)^{-1/2} 
\nonumber
\end{eqnarray}
for the constraints (\ref{constraintII0}),
The functions $g_1(p)$ and $g_2(p)$ can be expressed by 
elliptic integrals, which converge for $p>0$ and $p<-1$.
Note that for $p<-1$ the  functions   $g_1(p)$ and $g_2(p)$ are imaginary.

In addition the cap lengths have to fulfill  
$L_{\rm cap,1}+L_{\rm cap,2} + L_{st} =L$.
A sixth equation arises 
because the two caps can exchange 
 length while the adhered length $L_{st}$ stays fixed.
This leads to the  additional  transversality condition
that the contact curvatures of the two caps 
have to be equal. This condition 
enforces that both caps are {\em identical}:
Because at all four 
contact points $\sin \theta ({s_{co}}) = 0$, the Euler Lagrange 
equations (\ref{ELII0}) lead to equal integration
constants $c_i$ are equal for both caps, $c\equiv c_{1}=c_{2}$ 
or $\nu_i = c/p_i$.
Together with the two 
equations for the constraints  (\ref{astII0})
it follows that 
$p\equiv p_1=p_2$. Therefore, both caps are identical and 
also have  the same size
$L_{\rm cap} = L_{\rm cap,1}= L_{\rm cap,2}$. 
As a result we are left with two parameters $p$ and $\nu$ 
to be determined with  $L_{\rm cap}$ fixed by 
 $2L_{\rm cap} = L-L_{st}$ for  a prescribed adhered length.

Instead of explicitly solving for $p$ and $\nu$, 
 we express all quantities of interest 
as functions of $p$ using (\ref{LcapII0}) and (\ref{astII0}):
\begin{eqnarray}
  \bar{L}_{\rm cap} &=& \frac{L_{\rm cap}}{a_{st}}   =
      \frac{g_1(p)}{g_2(p)}
\label{LcappII0} \\
\bar{L}_{st} &=&   \frac{L_{st}}{a_{st}}  =  \bar{L} - 
          \frac{2g_1(p)}{g_2(p)}
\label{LstpII0}\\
  \frac{\nu a_{st}^2}{\kappa} &=& \frac{1}{2} g_2(p)^2
\label{nupII0}
\end{eqnarray}
The bending energy of the caps becomes 
\begin{eqnarray}
    E_{b} &=& 2E_{\rm cap}=   \left( {2\kappa}{\nu} \right)^{1/2}g_3(p)  
\label{EbII0}\\
   g_3(p) &\equiv &  \int_0^\pi d\theta (p+\sin\theta)^{1/2} 
\nonumber
\end{eqnarray}
where  the function $g_3(p)$ can be expressed by elliptic integrals. 
For $p<-1$ all  three  functions   $g_i(p)$  are imaginary but 
physical quantities remain real-valued. 
 It follows that 
\begin{equation}
  \bar{E}_b =  \frac{E_b a_{st}}{\kappa}  =  g_2(p) g_3(p)
\label{EbpII0}
\end{equation}
which gives together with (\ref{LstpII0}) a parametric representation of 
$E_b(L_{st})$  using the parameter $p$ in the range  $p>0$ and $p<-1$. 
The corresponding curve is shown in Fig.\ \ref{fig:landscape}(a) as blue line.

For $p\approx -1$ both $g_1(p)$ and $g_2(p)$ diverge  on the negative 
imaginary axis, whereas $g_3(-1) \approx -i 4 (\sqrt{2}-1)$ such that 
$\bar{L}_{st} \approx \bar{L} -2$ or $L_{\rm cap} \approx a_{st}$ 
 corresponding to the 
limit of maximal adhered length and a ring configuration approaching a 
rectangular shape. Accordingly $E_b$ diverges in this limit,
\begin{equation}
\bar{E}_b = 2\bar{E}_{b,{\rm cap}} \approx 32(\sqrt2 -1)^2
       (\bar{L}-2-\bar{L}_{st})^{-1}.
\end{equation}
For $p\gg 1$ (and similarly for $p\ll -1$),  we have 
$g_1(p) \approx \pi p^{-1/2}$,  $g_2(p) \approx \pi p^{-1/2}$, and $g_3(p)
\approx  \pi p^{1/2}$ such that  $\bar{L}_{st} \approx \bar{L} -\pi$ 
or $L_{\rm cap} \approx \pi a_{st}/2$  corresponding to exactly 
 semicircular caps with 
radius $R_{co} = a_{st}/2$. The bending energy is 
\begin{equation}
   \bar{E}_b = 2\bar{E}_{\rm cap} \approx 2\pi.
\end{equation}
For $p\approx 0$, finally, the caps become very elongated with large
contact radius $R_{co}$ and $g_1(0) \approx
5.24$, $g_2(0) = g_3(0) \approx 2.40$ such that the adhered length assumes 
its minimal value  $\bar{L}_{st} \approx \bar{L} -4.38$ or 
$\bar{L}_{\rm cap} \approx 2.19$, 
and the bending energy becomes 
\begin{equation}
\bar{E}_b = 2\bar{E}_{b,{\rm cap}} \approx 5.74.
\end{equation}

If the constraint of fixed adhered length $L_{st}$ is lifted 
the transversality 
condition of contact curvature at the contact points  gives
$c= |W_{st}|$  or 
\begin{equation}
   p = \frac{|W_{st}|}{\nu}
\end{equation}
which leads to  
\begin{equation}
    |w_{st}| = \frac{|W_{st}|a_{st}^2}{\kappa} = \frac{1}{2}p g_2^2(p)
\label{wpII0}
\end{equation}
Together with (\ref{EbpII0}) this gives  a parametric 
representation of $\bar{E}_{\rm tot}(|w_{st}|)$ using the parameter $p$ 
in the range  $p>0$ and $p<-1$. 
The corresponding curve is shown in Figs.\ \ref{eminv}(a,b) as blue line.

The limiting case $p\approx -1$ with $\bar{L}_{st} \approx \bar{L} -2$ 
corresponds to strong adhesion with $|w_{st}|\gg 2$ and 
\begin{equation}
  \bar{E}_{\rm tot,II_0} \approx 8\sqrt{2}(\sqrt{2}-1) |w_{st}|^{1/2}
         -|w_{st}| (\bar{L}-2).
\label{Etotapp1II0}
\end{equation} 
For  $|p|\gg 1$ we find  intermediate adhesion strengths 
$|w_{st}|\approx 2$ and 
\begin{equation}
  \bar{E}_{\rm tot,II_0} \approx 2\pi -|w_{st}| (\bar{L}-\pi).
\label{Etotapp3II0}
\end{equation}
The limiting case 
 $p\approx  0$ with $\bar{L}_{st} \approx \bar{L} -4.38$
corresponds to weak adhesion with $|w_{st}|\ll 1$ and 
\begin{equation}
  \bar{E}_{\rm tot,II_0} \approx 5.74 -|w_{st}| (\bar{L}-4.38)
  -1.31 |w_{st}|^{3/2}.
\label{Etotapp2II0}
\end{equation} 
The asymptotic estimate  (\ref{Etotapp2II0}) is
shown in Fig.\ \ref{eminv}(a) as dashed line.

\subsection{Shape ${\rm II_2}$}

Shape ${\rm II}_2$ consists of two bulges of lengths 
 $L_{\rm bul,1}$ and 
$L_{\rm bul,2}$  and two adhered segments 
with total  length $L_{st}$ and  $L_{\rm bul,1}+L_{\rm bul,2} + L_{st} =L$. 
As opposed to the caps of shape ${\rm II_0}$ bulges have 
no reflection symmetry. 
Also the bulges of shape ${\rm II_2}$ can exchange length
with the adhered length $L_{st}$ fixed. As for the caps, this 
leads to a transversality constraint that  curvatures at contact points 
connected by an adhered segment have to be equal. 
Using analogous arguments as for caps in shape ${\rm II_0}$ 
this leads  to the conclusion that both bulges must be 
identical in size, $L_{\rm bul} = L_{\rm bul,1}= L_{\rm bul,2}$. The 
bulge length is fixed by 
 $2L_{\rm bul} = L-L_{st}$  for  a prescribed adhered length.

There are two energetically degenerate configurations 
of the two  bulges in shape ${\rm II_2}$: 
An arrangement with reflection symmetry
with respect to the $y$-axis  and both bulges on the same side of the stripe, 
  and an antisymmetric arrangement with both bulges on opposite sides.

\begin{figure*}
\begin{center}
\includegraphics[width=0.98\textwidth]{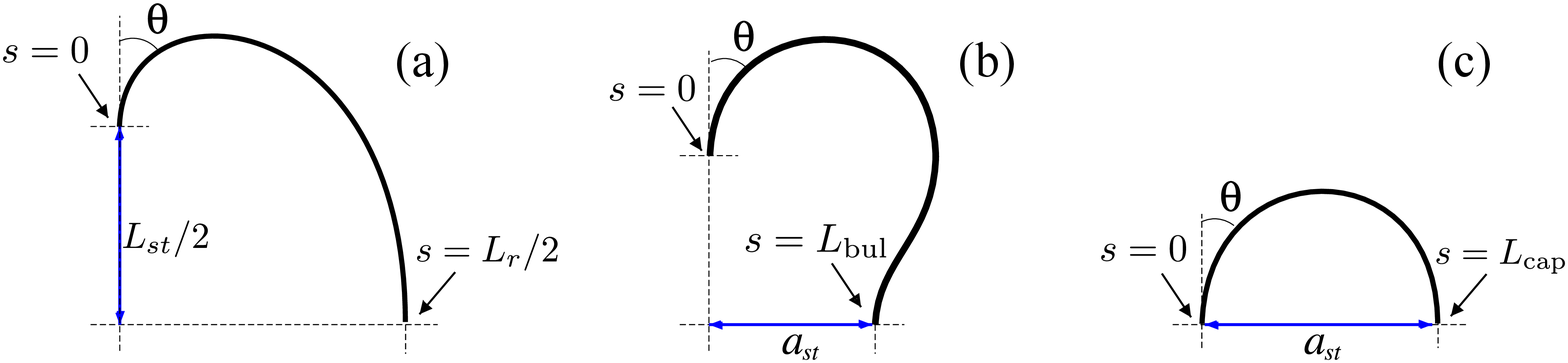}
\caption{
\label{fig:sketches} 
Shapes ${\rm I}$,  ${\rm II}_0$, ${\rm II}_1$ and ${\rm II}_2$ in
Fig. \ref{types}(a) 
 can be divided into straight (adsorbed) segments of total length $L_{st}$ 
and curved (desorbed) segments of the three types shown in this figure. 
Configuration ${\rm I}$ consists of two  curved segments as shown in (a).
Shape ${\rm II}_0$, consists of two straight segment and two 
unbulged caps as shown in (c).
 Shape  ${\rm II}_1$ consists of two straight segments and 
one bulged cap as shown in (b) and  on unbulged cap as shown in (c).
Shape ${\rm II}_2$ consists of two straight segments and 
two bulged caps as shown in (b). }
\end{center}
\end{figure*}

The ring closure constraint for the coordinate 
perpendicular to the groove is 
\begin{equation}
   \int_{L_{\rm bul}}ds\sin\theta(s)  -a_{st} = 0,
\label{constraintII2}
\end{equation}
which we associate with a Lagrange multiplier $\nu$. 
The Euler Lagrange equations and their first integral are identical 
to eqs.\ (\ref{EL1II0}) and (\ref{ELII0}) for the shape ${\rm II}_0$,
the integration constant $c$ also defines a parameter $p\equiv c/\nu$. 
The two unknown parameters $p$ and $\nu$ are determined 
by the two equations 
\begin{eqnarray}
   2L_{\rm bul} &=& L-L_{st} = \left( \frac{2\kappa}{\nu} \right)^{1/2}h_1(p)  
\label{LbulII2}\\
   h_1(p) &\equiv & \left( \int_0^\pi +2\int_\pi^{\theta_{\rm inf}} \right)
     d\theta (p+\sin\theta)^{-1/2} 
\nonumber
\end{eqnarray}
for the bulge  length and 
\begin{eqnarray}
   a_{st} &=& 2 \left( \frac{\kappa}{2\nu} \right)^{1/2}h_2(p)  
\label{astII2}\\
   h_2(p) &\equiv & 
  - \left( \int_0^\pi +2\int_\pi^{\theta_{\rm inf}} \right) 
 d\theta\cos\theta (p+\sin\theta)^{-1/2} 
\nonumber
\end{eqnarray}
for the constraint (\ref{constraintII2}).
Here, $\theta_{\rm inf}$ is the tangent angle in the inflection point of 
the bulge configuration, see Fig.\ \ref{fig:sketches}(c).
It is determined from $\partial_s\theta_{\rm inf} = 0$ which gives 
$\theta_{\rm inf} = \arcsin(-p)$ with $\pi< \theta_{\rm inf} < 3\pi/2$, which 
restricts $p$ to $0<p<1$.  
Also the functions $h_1(p)$ and $h_2(p)$ can be expressed by 
elliptic integrals. The function $h_2(p)$  becomes negative for 
$p>p_\infty\approx 0.652$, which restricts $p$ to $0<p<p_\infty$.

We express all quantities of interest 
 as functions of $p$ using (\ref{LbulII2}) and (\ref{astII2}):
\begin{eqnarray}
  \bar{L}_{\rm bul} &=& \frac{L_{\rm bul}}{a_{st}}   =
      \frac{h_1(p)}{h_2(p)}
\label{LbulpII2}\\
\bar{L}_{st} &=&   \frac{L_{st}}{a_{st}}  =  \bar{L} - 
          \frac{2h_1(p)}{h_2(p)}
\label{LstpII2}\\
  \frac{\nu a_{st}^2}{\kappa} &=& \frac{1}{2} h_2(p)^2
\label{nupII2}
\end{eqnarray}
According to (\ref{LbulpII2}) the length of the bulge diverges 
for $p\approx p_\infty$ as $\bar{L}_{\rm bul}\approx 
 5.72/(p_\infty-p)$.
Therefore, 
we can determine  a  $\bar{L}$-dependent value $p_L(\bar{L})<p_\infty$ 
such that 
$\bar{L}_{st}<0$ for $p>p_L(\bar{L})$, which sets 
the range $0<p<p_L(\bar{L})$ 
of accessible bulged states for a ring of finite 
length. For very large $\bar{L}$, $p_L(\bar{L}) \approx p_\infty$. 
For $\bar{L}=20$ as in 
Figs.\ \ref{fig:landscape} and \ref{eminv}, we find 
$p_L\approx 0.53$.

The bending energy can be rewritten as 
\begin{eqnarray}
   E_{b} &=& 2E_{b, {\rm bul}}=  \left( {2\kappa}{\nu} \right)^{1/2}h_3(p)  
\label{EbII2}\\
   h_3(p) &\equiv &  \left( \int_0^\pi +2\int_\pi^{\theta_{\rm inf}} \right) 
     d\theta (p+\sin\theta)^{1/2} 
\nonumber
\end{eqnarray}
where  also the function $h_3(p)$ can be expressed by elliptic integrals. 
 It follows that 
\begin{equation}
  \bar{E}_b =  \frac{E_b a_{st}}{\kappa}  =  h_2(p) h_3(p)
\label{EbpII2}
\end{equation}
which gives together with (\ref{LstpII2}) a parametric representation of 
$E_b(L_{st})$ using the parameter $p$ in the range  
$0<p<p_L(\bar{L})$ of accessible parameters $p$. 
The corresponding curve is shown in Fig.\ \ref{fig:landscape}(a) as 
green line.

For $p\approx 0$ the bulge of shape ${\rm II_2}$ approaches the 
 maximally elongated cap of shape ${\rm II_0}$,
 and the adhered length approaches the above result 
 $\bar{L}_{st} \approx \bar{L} -4.38$ or 
$\bar{L}_{\rm bul}=\bar{L}_{\rm cap} \approx 2.19$ with a 
bending energy $\bar{E}_b = 2\bar{E}_{b,{\rm bul}} \approx 5.74$.

If the constraint of fixed adhered length $L_{st}$ is lifted 
the condition of contact curvature at the 
contact points  gives $c= |W_{st}|$ or 
\begin{equation}
   p = \frac{|W_{st}|}{\nu}
\end{equation}
which leads to  
\begin{equation}
    |w_{st}|  = \frac{1}{2}p h_2^2(p)
\label{wpII2}
\end{equation}
This relation gives 
 a maximal value $|w_{st}| =w_{\rm max}\approx  0.35$ which is realized for 
$p=p_{\rm max} \approx 0.25$. For $|w_{st}|> w_{\rm max}$, 
eq.\ (\ref{wpII2}) has no solution because  bulged 
shapes ${\rm II_2}$ are no longer metastable states and are 
unstable with respect to  transitions into shape 
${\rm II_0}$. 
For $|w_{st}|< w_{\rm max}$, there are two solutions $p$ to eq.\
 (\ref{wpII2}). The solution branch with $p>p_{\rm max}$ corresponds 
to the  local energy minimum representing shape ${\rm II_2}$ 
whereas the solution branch with  $p<p_{\rm max}$
corresponds to a local maximum of the total energy and, thus, 
represents a possible transition state ${\rm II_2^*}$ 
for shape transitions into 
 shapes ${\rm II_1}$ or ${\rm II_0}$. 
This maximum corresponds to a shape with two identical {\em small}
 bulges which are  unstable with respect to shrinking 
to zero size to a shape ${\rm II_0}$ or to expanding to its 
equilibrium size in state ${\rm II_2}$. 
Eqs.\ (\ref{wpII2}) and 
 (\ref{EbpII2}) give  a parametric 
representation of  $\bar{E}_{\rm tot}(|w_{st}|)$ for a metastable 
shape ${\rm II_2}$ using the parameter $p$ in 
the  range $p_{\rm max}<p<p_L(\bar{L})$.
 The value  $p=p_L(\bar{L})$ corresponds to a minimal  value 
$|w_{st}| =w_{\rm min}(\bar{L})$, which is 
$\bar{L}$-dependent. For  $|w_{st}|> w_{\rm min}$,
bulged shapes  ${\rm II_2}$ become  
unstable with respect to  transitions into shape 
${\rm I}$ because bulged become so large that 
the adhered length $\bar{L}_{st}$ vanishes.
For $\bar{L}=20$ we find $w_{\rm min} \approx 0.10$.   
The corresponding curve is shown in Figs.\ \ref{eminv}(a,b) 
for $\bar{L}=20$ as green line in the corresponding 
range $w_{\rm min} < |w_{st}|< w_{\rm max}$. 
For the range $0<p<p_{\rm max}$ the parametric representation 
gives  the additional 
branch of transition states ${\rm II_2^*}$ 
 shown in Fig.\ \ref{eminv}(b) as green dashed line
in the corresponding range $0 < |w_{st}|< w_{\rm max}$.

For $p\approx p_\infty$ the bulge length diverges. 
This limit corresponds to weak adhesion 
with $|w_{st}|\ll 1$. Expanding  
 the functions $h_i(p)$ around  $p\approx p_\infty$ we find  
\begin{eqnarray}
   \bar{L}_{\rm bul}&\approx&  |w_{st}|^{-1/2}
       \left(3.75 -1.53 |w_{st}|^{1/2}- 0.54|w_{st}|\right)
\nonumber\\
   \bar{E}_{b,{\rm bul}} &\approx& |w_{st}|^{1/2} 
      \left( 3.75 +0.18 |w_{st}| \right)
\nonumber\\
  \bar{E}_{\rm tot,II_2} &\approx&  15.00 |w_{st}|^{1/2} 
      -|w_{st}|(\bar{L} +3.07)- 0.72 |w_{st}|^{3/2}.
\nonumber\\
\label{Etotapp1II2}
\end{eqnarray} 
This asymptotic estimate  is shown in Fig.\ \ref{eminv}(a) 
for $\bar{L}=20$ as dashed  line
in the accessible range $p_{\rm
  max}<p<p_L(\bar{L})$ corresponding to $0.10 < |w_{st}|< w_{\rm max}$.

We can also define an energy $\Delta \bar{E}_{\rm bul}$ for creating a 
bulge starting from the shape ${\rm II_0}$. This energy includes 
the bending energy gain of a bulge as compared to a cap as well as 
the adhesion energy cost from desorbing additional length,
\begin{eqnarray}
   \Delta \bar{E}_{\rm bul} &=& (E_{\rm tot,II_2}-E_{\rm tot,II_0})/2 
\nonumber\\
    &\approx&   -2.87+  7.50 |w_{st}|^{1/2}  -  3.72 |w_{st}|
  +0.30|w_{st}|^{3/2}
\nonumber\\ 
  \label{DeltaEbulapp}
\end{eqnarray}
where the last approximation holds for $|w_{st}|\ll 1$.

\subsection{Shape ${\rm II_1}$}

Shape ${\rm II}_1$ consists of one bulge 
 of length $L_{\rm bul}$, one cap of length $L_{\rm cap}$,  
and two adhered segments 
with total length $L_{st}$ such that $L_{\rm bul} + L_{\rm cap} + L_{st} =L$. 
We have to consider bulge and cap separately and apply two 
constraints 
\begin{eqnarray}
   \int_{L_{\rm bul}}\sin\theta(s)  -a_{st} &=& 0
\nonumber\\
    \int_{L_{\rm cap}}\sin\theta(s)  -a_{st} &=& 0
\label{constraintII1}
\end{eqnarray}
which we associate with two Lagrange multipliers $\nu_{\rm bul}$ and
 $\nu_{\rm cap}$. 
The Euler Lagrange equations and their first integral are identical
to eqs.\ (\ref{EL1II0}) and (\ref{ELII0}) for the shapes ${\rm II}_0$ 
and ${\rm II_2}$.
Because the adhered length between bulge and cap can be adjusted in 
shape ${\rm II_1}$ we have an additional transversality condition
that the contact curvatures have to be equal. Because at all four 
contact points $\sin \theta ({s_{co}}) = 0$, the Euler Lagrange 
equations (\ref{ELII0}) lead to the equivalent 
condition that the integration 
constants are equal for bulge and cap, $c= c_{\rm bul}=c_{\rm cap}$. 
We introduce two corresponding parameters 
$p_1$ and $p_2$ such that 
\begin{equation}
c= p_1 \nu_{\rm bul} = p_2 \nu_{\rm cap}.
\label{cII1}
\end{equation}
This relation together with $L_{\rm bul} + L_{\rm cap} + L_{st} =L$ and 
the   four equations for cap and bulge length and  constraints, 
\begin{eqnarray}
   L_{\rm bul} &=&  \left( \frac{\kappa}{2\nu_{\rm bul}} \right)^{1/2}h_1(p_1) 
\label{LbulII1}\\
   L_{\rm cap} &=&\left( \frac{\kappa}{2\nu_{\rm cap}} \right)^{1/2}g_1(p_2)  
\label{LcapII1}\\
  a_{st} &=& 2 \left( \frac{\kappa}{2\nu_{\rm bul}} \right)^{1/2}h_2(p_1)  
\label{astII1a}\\
   a_{st} &=& 2 \left( \frac{\kappa}{2\nu_{\rm cap}} \right)^{1/2}g_2(p_2)  
\label{astII1b}
\end{eqnarray}
give six equations for the six parameters $L_{\rm bul}$, $L_{\rm cap}$, 
$p_1$, $p_2$, $\nu_{\rm bul}$, and $\nu_{\rm cap}$.

From these equations we find 
\begin{eqnarray}
&&\bar{L}_{\rm bul} =   \frac{h_1(p_1)}{h_2(p_1)}
~~,~~ \bar{L}_{\rm cap} =   \frac{g_1(p_2)}{g_2(p_2)}
\nonumber\\
&&  p_1 h_2^2(p_1) =  p_2 g_2^2(p_2)
\label{p1p2}
\end{eqnarray}
Using the last equation we can solve numerically for $p_2$ 
which allows us to express the adhered length
 $L_{st}= L -L_{\rm bul}-L_{\rm cap}$ 
parametrically as a function of $p_1$.
It follows that the adhered length of shape ${\rm II_1}$ 
with one bulge and one cap 
for given $p_1$ and  corresponding $p_2$ is 
 the mean value of the adhered lengths of shapes ${\rm II_2}$ 
with two bulges for $p=p_1$
and ${\rm II_0}$  two caps for $p=p_2$,
\begin{equation}
   L_{st,II_1}(p_1) = \frac{1}{2}
       \left(L_{st,II_2}(p_1)+L_{st,II_0}(p_2)\right).
\label{LstpII1}
\end{equation}
Therefore, 
as for the shape ${\rm II_2}$, the length of the bulge diverges for 
$p_1\approx p_\infty$. 
There exists a  
$\bar{L}$-dependent value $p_{1,L}(\bar{L})<p_\infty$ 
such that the adhered length on one of the stripe edges 
shrinks to zero, which sets 
the range $0<p<p_{1,L}(\bar{L})$ 
of accessible bulged states for a ring of finite 
length. For very large $\bar{L}$, $p_{1,L}(\bar{L}) \approx p_\infty$. 
For $\bar{L}=20$ as in 
Figs.\ \ref{fig:landscape} and \ref{eminv}, we find 
$p_{1,L}\approx 0.55$.

Furthermore the bending energy becomes 
\begin{eqnarray}
    E_{b} &=& E_{\rm bul}+E_{\rm cap} 
\nonumber\\
  &=&    
    \left( {2\kappa}{\nu_{\rm bul}} \right)^{1/2}h_3(p_1)  +
   \left( {2\kappa}{\nu_{\rm cap}} \right)^{1/2}g_3(p_2).
\label{EbII1}
\end{eqnarray}
Using (\ref{p1p2}) we can express also the bending energy parametrically 
as a function of $p_1$.
The bending energy in  shape ${\rm II_1}$ is given by 
 the mean value of the bending energies of shapes ${\rm II_2}$ 
with two bulges for $p=p_1$
and ${\rm II_0}$  two caps for $p=p_2$,
\begin{equation}
   E_{b,{\rm II_1}}(p_1) =\frac{1}{2}
       \left(E_{b,{\rm II_2}}(p_1)+E_{b,{\rm II_0}}(p_2)\right).
\label{EbII1a}
\end{equation}
Together with the parametric result for $L_{st}$ 
we obtain  a parametric representation of 
$E_b(L_{st})$ using the parameter $p_1$ in the range  
$0<p_1<p_{1,L}(\bar{L})$ of accessible parameters $p_1$. 
The corresponding curve is shown in Fig.\ \ref{fig:landscape}(a) as 
violet line.

If the constraint of fixed adhered length $L_{st}$ is lifted 
the condition of contact curvature at the contact points  gives 
$c= |W_{st}|$ or 
\begin{equation}
   p_1 = \frac{|W_{st}|}{\nu_{\rm bul}}
~~\mbox{and}~~
   p_2 = \frac{|W_{st}|}{\nu_{\rm cap}}
\end{equation}
which leads to  
\begin{equation}
    |w_{st}|  = \frac{1}{2}p_1 h_2^2(p_1) = \frac{1}{2}p_2 g_2^2(p_2)
\label{wpII1}
\end{equation}
which is equivalent to the above relation 
 (\ref{p1p2}) between $p_1$ and $p_2$.
It follows that for a given value of $w_{st}$ the condition of 
the same contact curvature at all four contact points automatically 
leads to values $p_1$ and $p_2$ satisfying (\ref{p1p2}). 
According to (\ref{LstpII1}) and (\ref{EbII1a}), we conclude that 
the total energy of shape ${\rm II_1}$ is exactly 
the mean value of the total energies of shape ${\rm II_0}$ and 
shape ${\rm II_2}$,
\begin{eqnarray}
   E_{\rm tot,II_1} &=&\frac{1}{2}
       \left(E_{\rm tot,II_2}+E_{\rm tot,II_0}\right)
\nonumber\\
   &=& E_{\rm tot,II_0} + \Delta E_{\rm bul}
\label{EtotII1}
\end{eqnarray}
for the same value of the adhesion strength $w_{st}$. 
The shape ${\rm II_1}$ exists for 
 $p_{\rm max}<p_1<p_{1,L}(\bar{L})$
corresponding to $w_{\rm min}(\bar{L}) < |w_{st}|< w_{\rm max}$ 
with $w_{\rm min} \approx 0.07$ for  $\bar{L}=20$.
The resulting  curve is shown in Figs.\ \ref{eminv}(a,b)
as violet solid line.  A corresponding 
 asymptotic estimate  is shown in Fig.\ \ref{eminv}(a) 
 as dashed  line.
There is also a shape  ${\rm II_1^*}$ 
corresponding to local maximum, which plays the role of a possible transition 
state,  for which we find the analogous result 
\begin{equation}
   E_{\rm tot,{II_1^*}} =\frac{1}{2}
       \left(E_{\rm tot,{II_2^*}}+E_{\rm tot,II_0}\right).
\label{EtotII1s}
\end{equation}
This maximum corresponds to a shape with a   {\em small}
 bulge which is  unstable with respect to shrinking 
to zero size to a shape ${\rm II_0}$ or to expanding to its 
equilibrium size in state ${\rm II_1}$. 
The shape ${\rm II_1^*}$  exists for $0<p_1<p_{\rm max}$ 
corresponding to  $0 < |w_{st}|< w_{\rm max}$
and is shown in Fig.\ \ref{eminv}(b) as violet dashed line.

\subsection{Other bulged shapes}

In principle, there exists also  metastable states 
with bulges that extend to {\em both} sides of the groove on the 
same curved segment. 
Analogously to the results for shape ${\rm II_2}$, we can calculate 
the length $L_{\rm bul}^*$ of such a bulge and  
the energy of a state ${\rm II_4}$ with two such bulges 
for fixed total adhered length $L_{st}= L - 2L_{\rm bul}^*$ from 
\begin{eqnarray}
   2L_{\rm bul}^* &=& L-L_{st} = \left( \frac{2\kappa}{\nu} \right)^{1/2}k_1(p)  
\label{LbulII4}\\
   k_1(p) &\equiv & \left( \int_0^\pi +4\int_\pi^{\theta_{\rm inf}} \right)
     d\theta (p+\sin\theta)^{-1/2} 
\nonumber\\
   a_{st} &=& 2 \left( \frac{\kappa}{2\nu} \right)^{1/2}h_2(p)  
\label{astII4}\\
   k_2(p) &\equiv & 
  - \left( \int_0^\pi +4\int_\pi^{\theta_{\rm inf}} \right) 
 d\theta\cos\theta (p+\sin\theta)^{-1/2} 
\label{k2p}
\end{eqnarray}
resulting in 
\begin{eqnarray}
  \bar{L}_{\rm bul}^* &=& 
      \frac{k_1(p)}{k_2(p)}
\label{LbulpII4}\\
\bar{L}_{st} &=&   \bar{L} - 
          \frac{2k_1(p)}{k_2(p)}
\label{LstpII4}\\
  \frac{\nu a_{st}^2}{\kappa} &=& \frac{1}{2} k_2(p)^2
\label{nupII4}
\end{eqnarray}
and 
\begin{eqnarray}
   E_{b} &=& 2E_{\rm bul}^*=  \left( {2\kappa}{\nu} \right)^{1/2}k_3(p)  
\nonumber \\
   \bar{E}_b &=&  k_2(p) k_3(p)
\label{EbpII4}\\
   k_3(p) &\equiv &  \left( \int_0^\pi +\int_\pi^{\theta_{\rm inf}} \right) 
     d\theta (p+\sin\theta)^{1/2} 
\nonumber
\end{eqnarray}
This  gives together with (\ref{LstpII4}) a parametric representation of 
$E_b(L_{st})$ using the parameter $p$ in the range  
$0<p<p_{4,L}(\bar{L})$ of accessible parameters $p$, where 
$p_{4,L}(\bar{L})$ is determined by the condition that 
$\bar{L}_{st}<0$ for $p>p_{L,4}(\bar{L})$.

The resulting parametric representation of $E_b(L_{st})$ shows that
$E_{b,{\rm II_4}}(L_{st})> E_{b,{\rm II_2}}(L_{st})$ for {\em all} possible 
values of $L_{st}$. Therefore, the bending energies of 
 bulges which extend to both sides 
of the groove are always higher in bending energy for the 
same adhered length $L_{st}$.

\section{Analytical energy minimization for ring condensation}
\label{appendix2}

In this appendix we derive  exact analytical result for the 
metastable racquet shape of a ring in the presence of a
polymer-polymer attraction $W_{\rm con}<0$  per contact length. 

The racquet shape consists of two round bulges, which assume the 
same length $L_{\rm bul}$ in equilibrium according to analogous arguments  
as for rings on the topographical stripe, and two 
adhering straight segments with total length $L_{\rm con}$, which are in contact 
 with contact energy
$-|W_{\rm con}|L_{\rm con}/2$. 

The ring closure constraint for the coordinate
perpendicular to the adhering segments  is 
\begin{equation}
   \int_{L_{\rm bul}}ds\sin\theta(s)  = 0,
\label{constraintrac}
\end{equation}
which we associate with a Lagrange multiplier $\nu$. 
The Euler Lagrange equations and their first integral are identical 
to eqs.\ (\ref{EL1II0}) and (\ref{ELII0}) for the shape ${\rm II}_0$,
the integration constant $c$ also defines a parameter $p\equiv c/\nu$. 

The bulges of the 
racquet shape can be treated  analogously  to the bulges 
of shape ${\rm II_4}$  for the topographical surface groove,  and we find
\begin{eqnarray}
   2L_{\rm bul} &=& L-L_{\rm con} = 
   \left( \frac{2\kappa}{\nu} \right)^{1/2}k_1(p).  
\label{Lbulrac}
\end{eqnarray}
The ring closure constraint (\ref{constraintrac}) gives 
\begin{eqnarray}
   0  &=& \left( \frac{\kappa}{2\nu} \right)^{1/2}k_2(p)  
\label{constrac}
\end{eqnarray}
such that $p=p_0\approx 0.46$ must be a zero of the elliptic 
function $k_2(p)$ defined in (\ref{k2p}). 
Eq.\ (\ref{Lbulrac}) 
with $p=p_0$ then gives the Lagrange multiplier $\nu$ as a function of the 
contact length $L_{\rm con}$. 
The bending energy is 
\begin{eqnarray}
   E_{b} &=&   \left( {2\kappa}{\nu} \right)^{1/2}k_3(p_0)  
\label{Ebrac}
\end{eqnarray}
as a function of $\nu$.

If the constraint of fixed contact length $L_{\rm con}$ is lifted 
the condition of contact curvature at the 
contact points  gives $c= |W_{\rm con}|$ and, thus,  
determines the Lagrange multiplier $\nu  = |W_{\rm con}|/p_0$.
 Using this we obtain for  
 the contact length from eq.\ (\ref{Lbulrac}),
\begin{eqnarray} 
    \frac{L_{\rm con}}{L} &=& 
      1- 
   \left( \frac{\kappa}{|W_{\rm con}|L^2} \right)^{1/2}
       (2p_0)^{1/2}k_1(p_0).  
\label{Lconrac}
\end{eqnarray}
and for the bending energy 
\begin{eqnarray} 
    \frac{E_bL}{\kappa} &=& 
        \left(\frac{|W_{\rm con}|L^2}{\kappa} \right)^{1/2}  
         \frac{2^{1/2}k_3(p_0)}{p_0^{1/2}}.  
\label{Ebrac2}
\end{eqnarray}
The total energy $E_{\rm tot,rac} = E_b -|W_{\rm con}|L_{\rm con}/2$ 
is obtained as
\begin{eqnarray}
   \frac{E_{\rm tot,rac}L}{\kappa}\!\! &=&  
      \!\! \left(\frac{|W_{\rm con}|L^2}{\kappa}\right)^{1/2}\!  (2p_0)^{1/2}
       \left( \frac{k_3(p_0)}{p_0}+ \frac{k_1(p_0)}{2} \right)
      \nonumber\\
    && ~~~ 
       -  \frac{1}{2} \frac{|W_{\rm con}|L^2}{\kappa}
\label{Etotapprac}
\end{eqnarray}
with $ (2p_0)^{1/2}
       \left( k_3(p_0)/p_0+ k_1(p_0)/2 \right)\approx 12.85$.
This result holds for nonzero contact length $L_{\rm con}>0$ or 
 $|W_{\rm con}|L^2/\kappa> 2p_0 k_1^2(p_0)\approx 73.33$.




\footnotesize{
\bibliography{polymerrings_SoftMatter} 
\bibliographystyle{rsc} 
}

\end{document}